# A polynomial time algorithm for 3SAT

Lizhi, Du*

College of computer science and technology, Wuhan university of science and technology, Wuhan 430065, email: dlz95@sohu.com

**Abstract**. By creating some new concepts and methods: checking tree, long unit path, direct contradiction unit pair, indirect contradiction unit pair, additional contradiction unit pair, 2-unit layer and 3-unit layer, redundant units, and destroying parallel pairs, we successfully transform solving a 3SAT problem to solving 2SAT problems in polynomial time. Thus we proved that NP=P. The key steps are: we develop the good checking tree, the standard checking tree, the way to get the final leaf with its good units. Then by the final leaf combination property and the correct chosen units property, the time complexity is polynomial.

## 1 INTRODUCTION

Some experts like to assert that NP is not equal to P, but many experts had asserted something in history, and later these assertions have been broken by new achievements. Also there are some famous scientists agree that NP=P. Hilbert, a great mathematician of the twentieth century, has a famous saying: we must know; we will know. It can be seen that Hilbert essentially agreed that NP equals P, though in his time, the concept NPC did not appear. Many mathematical problems in human history, including Hilbert's famous 23 mathematical problems, are constantly being solved. These facts strongly imply that NP equals P. This is because that if NP is not equal to P, a human being brain cannot solve a lot of hard problems in polynomial time[3].

From the heuristic point of view, each NPC problem can be reduced to any other NPC problem in polynomial time. That is to say, every distance between two NPC problems is polynomial. The fact itself strongly shows that NP problems have a unified solution law and difficulty, and its solution difficulty should be polynomial order of magnitude. The difference of an attribute value between any group of individuals in the objective world is usually in the same order of magnitude as the absolute value of an individual attribute. For example, one adult weighs in 100 pounds, and the difference between a very fat man and a very thin man is also in 100 pounds. Similarly, the weight of an ant is in gram, and the difference between a big ant and a small ant is also in gram. Etc. Of course, these are not strictly proven conclusions.

We develop a polynomial time algorithm for 3SAT. This paper contains key three steps. Key step 1: get a good checking tree which has these properties: each unit is in at least one long path; for any two units, if they are a contradiction pair, no any one long path to contain them and if they are not a contradiction pair, there is at least one long path to contain them; we know all contradiction pairs. Key step 2: after some units were deleted from a good checking tree, it was destroyed. We can repair it by a standard checking tree. Key step 3: by algorithm E, we construct 2SAT based on the standard checking tree. The extraordinary insight of this paper is: the strong function of the standard checking tree property and the destroying parallel pair can lead the 2SAT logic. **Note that the parallel part paths** over the first two units in each generation without intersection part paths between them is the key for the 2SAT logic. Based on these, if there are long paths, our goal is to get the last real long path. For each time to do algorithm E1, we get a final leaf and its good units. At last we can get one final leaf whose good units are all in the last real long path. We call units in the last real long path correct units. By the standard checking tree property, at last, we can get a combination of only correct final leaves with their correct good units in polynomial time. For this result, the key is the final leaf combination property, especially **the correct chosen units property**. Also we update contradiction pairs for the whole clauses. So this algorithm is not based on "local consistency", but on the whole and we only calculate each unit $O(n)$ times, not exponential.

* Lizhi Du, College of computer science and technology, Wuhan university of science and technology. Email: dlz95@sohu.com

## 2. ALGORITHM AND PROOF

### 2.1 Foundation Method

A 3SAT contains *n* variables and *m* clauses. Each clause contains three literals. We call each of them a unit (**we call it a unit not a literal in order to separate the same one literal which is in different clauses**). So there are 3*m* units in all the clauses. Now we change this 3SAT to a path finding problem. There are *m*+1 cities: $c_0$, $c_1$, ... $c_m$. From $c_i$ to $c_{i+1}$ (*i*=0, 1, ... *m*-1), there are 3 different roads. We call each road a unit. So there are 3*m* units. Now we want to find a path from $c_0$ by $c_1$, ...$c_{m-1}$ to $c_m$. We call such a path a long path. There are $3^m$ different possible long paths. A long path contains *m* units. But a lot of two units have contradictions. Such two units cannot be in the same path. There are a lot of two units which cannot be in the same path. For any two units, we know they have or do not have contradictions (in 3SAT, a variable *x* and -*x* have contradictions). Now the question is: how to find a long path from $c_0$ to $c_m$?

If two units have contradictions, we say that one unit destroys the other one.

Suppose that from city $c_0$ to $c_1$, the 3 roads are $a_{11}$, $a_{12}$, $a_{13}$ (they are three literals in the first clause in 3SAT); from $c_1$ to $c_2$, they are $a_{21}$, $a_{22}$, $a_{23}$, ... and from $c_{m-1}$ to $c_m$, they are $a_{m1}$, $a_{m2}$, $a_{m3}$. All these are 3*m* units in 3SAT.

The checking tree contains three roots $a_{11}$, $a_{12}$, and $a_{13}$. The roots are in the **layer 1**. We say layer 1 is the **highest layer** or the first layer in the tree. $a_{21}$, $a_{22}$, $a_{23}$ are in layer 2 and then $a_{31}$, $a_{32}$, $a_{33}$ are in layer 3. Layer 1 is higher than layer 2, and so on.

There are *m* clauses and each clause contains three units. So in the checking tree, at last there are *m* layers and each layer contains at most three units. A layer is versus a clause. All units are 3*m*. We call a lower layer a descendant layer of a higher layer. Also the latter is an ancestor of the former.

**A checking tree** contains *i* layers. Each layer contains three possible units. Each time we add one layer's units to the checking tree one by one. We first add the first layer's three units $a_{11}$, $a_{12}$, $a_{13}$ one by one. Then add $a_{21}$, $a_{22}$, $a_{23}$. And so on. We call the layer we currently calculate the **current layer**. The current layer's three units are $a_{i1}$, $a_{i2}$, $a_{i3}$ ($i$>2). We call a path from any one unit of the current layer to any one unit of $a_{11}$, $a_{12}$, $a_{13}$ the **current long path** or **a long path**. Each layer has one unit in this path and any two units in this path do not have contradictions.

If no current long path contains a unit *u*, we call *u* an **utterly destroyed unit**. *u* or every other unit which is the same literal as *u* will always be an utterly destroyed unit later.

If two units $u_1$ and $u_2$ are the same variable, but one is positive and the other one is negative, we say $u_1$ **directly destroys** $u_2$ (also $u_2$ destroys $u_1$). If anyone current long path does not contain the pair of $u_1$ and $u_2$ (two units together), we say that $u_1$ **indirectly destroys** (if does not directly destroy) $u_2$. For both cases, we also say one destroys the other one. If $u_1$ destroys $u_2$, it will always destroy $u_2$ later.

Each time we add one layer and then calculate. We add and calculate each unit of this layer one by one. We call it **the new added unit**. For each layer, we have to calculate all possible new utterly destroyed units, and, for every unit pair, i.e., for any two units in different layers, we have to calculate that whether one indirectly destroys the other one. For a unit pair, if one destroys the other one, we call it a **contradiction unit pair** or **a contradiction pair**. For direct destroying, it is a **direct contradiction pair** and for indirect destroying, it is an **indirect contradiction pair**.

In the checking tree, for *k* consecutive layers, we take one unit from each layer. If they do not destroy each other, we call these *k* units a **part path over these *k* layers,** or a path. In any part path, any two units do not have contradictions. Any consecutive part of a long path is a **part path**.

So, a checking tree contains *i* layers. Each layer contains at most three units. If a unit is an utterly destroyed unit, we do not keep it in the checking tree. For the checking tree, we have to remember all utterly destroyed units and all indirect contradiction unit pairs.

Now the current layer is the *i*th layer. Its three units are $a_{i1}$, $a_{i2}$, $a_{i3}$. We have got all utterly destroyed units and all contradiction unit pairs. Now we add and calculate the *i+1*th layer. Its three units are $a_{i+11}$, $a_{i+12}$, $a_{i+13}$. We add each one to the checking tree respectively. We first add the unit $a_{i+11}$ to the checking tree.

Now the old utterly destroyed units are still utterly destroyed units and the old contradiction unit pairs are also contradiction unit pairs. We have to calculate the new utterly destroyed units and the new indirect contradiction unit pairs after the $a_{i+11}$ is added. **Please note**: the contradiction pairs or utterly destroyed units are based on the literals. For example, if $u_1$ and $u_2$ are an indirect contradiction pair, and $u_3$ and $u_1$ are the same literals as well as $u_4$ and $u_2$ are the same literals, then $u_3$ and $u_4$ are also an indirect contradiction pair.

If the checking tree does not contain such a unit which is the same variable as $a_{i+11}$ but with different sign, i.e., if $a_{i+11}$ is the variable *v*, and no other unit is the variable -*v*, then there are no new utterly destroyed units and no new contradiction unit pairs (for $a_{i+11}$). If the checking tree contains the unit -*v*, then for a new long path which contains $a_{i+11}$, any unit whose variable is -*v* cannot be in this path.

If a layer only contains one unit (others are destroyed and are deleted), we call it a **1-unit layer**. In this way, we have **2-unit layer**, **3-unit layer**. We only consider 2-unit layers and 3-unit layers, so we call all layers which contain 2 or 1 unit 2-unit layers. Other layers are 3-unit layers.

For a checking tree with *k* layers, each layer contains one, two or three units. We know all direct contradiction pairs and indirect contradiction pairs. If there is no such a long path over the *k* layers which contains the unit pair $p_1$, then it is an indirect contradiction pairs (if not direct). So for any two units which are not a contradiction pair, there is at least one long path over the *k* layers which contains this unit pair and any one unit is in at least one long path. We call such a checking tree **a good checking tree** with *k* layers. The **good checking tree property**: each unit is in at least one long path over the *k* layers; for any two units, if they are a contradiction pair, no any one long path to contain them and if they are not a contradiction pair, there is at least one long path to contain them; we know all contradiction pairs. We delete some units from such a good checking tree and it still contains *k* layers, then we call it **a destroyed checking tree** with *k* layers.

We separate a destroyed checking tree into two parts. One contains **2-unit layers** which also include 1-unit layers, and the other one contains **3-unit layers**.

We call the two units in the same 2-unit layer **a parallel pair** and call one is the other one's **parallel pair mate**. Call the deleted unit in this layer **the parallel pair's third unit,** or **this layer's third unit**, though they may often change.

For a unit in 3-unit layers and a 2SAT path over the 2-unit layers (that is, a part path over all 2-unit layers), if this unit does not destroy any unit in this 2SAT path, we call the 2SAT path this unit's **one useful 2SAT path** or this unit has this useful 2SAT path and call any one unit in the 2SAT path this unit's **one useful unit**. Also we say that the former unit **has this useful unit**. We call any two units in its one useful 2SAT path this unit's **one useful unit pair**. Also we say that the former unit **has this useful unit pair, or useful pair**.

If one or two units in a 3-unit layer do not have any useful 2SAT paths, we delete these units and put this layer into 2-unit layers to recalculate. If in a 3-unit layer, no unit has any useful 2SAT paths, then we call this checking tree **a bad checking tree** and there is no path over all the *k* layers of the checking tree.

If a unit has a useful unit *i* but does not have the useful unit *j* which is *i*'s parallel pair mate (that is, *i* and *j* are in the same layer), we call *i* a **single useful unit** of this unit and this unit has a single useful unit *i*.

In a 3-unit layer, all three units' useful units together are **this layer's useful units,** and all three units' useful unit pairs together are **this layer's useful unit pairs**

For two units in 3-unit layers, if one unit has a single useful unit *i* and the other unit has a single useful unit *j* which is the parallel pair mate of *i*, that is, *i* and *j* are in the same layer, or, *i* and *j* are not in the same layer but they are a contradiction pair, then we set these two units (literals) in 3-unit layers (also in 2-unit layers) be a contradiction pair and call them an **additional contradiction pair**.

A part path or a long path means that any two units in it are not a contradiction pair (direct, indirect, additional).

We calculate a destroyed checking tree by algorithm 1 so that it becomes **a standard checking tree** if possible.

So from a destroyed checking tree to a standard checking tree, we may have some additional contradiction pairs and we also may delete some units.

Our goal is: judge whether there are some long paths (current long paths) over the *k* layers of a standard checking tree and when there are, get one. If a checking tree contains zero 3-unit layers, we can easily judge and get one if exists.

**The standard checking tree property**: it is based on a destroyed checking tree, some redundant units are deleted from this checking tree, some additional contradiction pairs are added, and both checking trees have the same number of layers; in 3-unit layers, each unit's each useful unit or useful unit pair is in at least one of its useful 2SAT paths over 2-unit layers; for each unit in 2-unit layers, if it is not in any one 2SAT path, delete it, so in 2-unit layers, each unit is in at least one 2SAT path over these layers; for each unit in a 3-unit layer, if it does not have any useful 2SAT path, delete it and move this layer to 2-unit layers; for each unit in 3-unit layers, if it destroys all units in another one 3-unit layer, then delete it and move its layer to the 2-unit layers; if all useful units of a unit $u_1$ in 3-unit layers do not contain a unit $u_2$ in 2-unit layers, then $u_1$ and $u_2$ are an additional contradiction pair (if they were not contradiction pair formerly); for any two units which are not contradiction pair in different 3-unit layers, if one's a single useful unit and the other one's a single useful unit are in the same layer, set these two units as an additional contradiction pair; for each unit *u*'s each useful unit or useful unit pair, if in another 3-unit layer, the units which do not destroy *u* do not have this useful unit or useful unit pair, *u* also has to lose this useful unit or useful unit pair and thus we have to recalculate. For a destroyed checking tree, we do algorithm 1 to let the standard checking tree property hold or get the result that the destroyed checking tree cannot become a standard checking tree.

In a word, t**he standard checking tree property is:** for each unit *u*'s each useful unit, in any other one 3-unit layer, there is at least one unit which does not destroy *u* and which also has this useful unit, and for each unit *u*'s each useful unit pair, in any other one 3-unit layer, there is at least one unit which does not destroy *u* and which also has this useful unit pair.

At any time when the standard checking tree property holds, we call each part path over the whole 2-unit layers a **useful 2SAT path** and call any two units in this part path **brother units**.

Next we discuss how to construct the checking tree and how to **calculate new contradiction pairs**:

We want to calculate new contradiction pairs after we add a new unit to the checking tree. Suppose the new added unit is $a_{i+11}$ and its variable is *v*. Before $a_{i+11}$ is added, the checking tree is a good checking tree. We call this good checking tree **the main checking tree**. For the main checking tree (good checking tree), we need to remember each layer's units, remember all contradiction pairs. Also remember all utterly destroyed units which have been deleted.

Now we copy the main checking tree to an empty tree and we will delete some units in this tree and then calculate it. We call it **the calculating main checking tree** (a temporary checking tree).

For two units *x* and *y*, the unit pair *x* and *y* is not a contradiction pair before $a_{i+11}$ is added. Now we want to know whether *x*, *y* can be in the same long path again after we delete all the -*v*. So we have to calculate whether *x* and *y* are a new contradiction pair after $a_{i+11}$ is added. For each unit pair which is not a contradiction pair before $a_{i+11}$ is added, we have to calculate one time in the same way.

For this temporary checking tree, we delete the layers which contain *x* and the layers which contain *y*. We find all layers which contain -*x* and delete all -*x*. We find all layers which contain -*y* and delete all -*y*. We find all layers which contain -*v* and delete all -*v*. So we delete all literals -*v*, -*x*, -*y* and call these literals **trouble units**. Then we call all layers each of which contains one or two units **2-unit layer**s. Before the trouble units are deleted, the checking tree is **a good checking tree** and after the trouble units are deleted, the checking tree is **a destroyed checking tree**. Then we calculate the destroyed checking tree by algorithm 1 to let it become a standard checking tree if possible. In the mean time, if all units in a layer are deleted, that is, we cannot get a standard checking tree, then there is no path over this checking tree and thus *x* destroys *y*.

Then call the algorithm E to calculate the standard checking tree, and we can determine whether *x* and *y* do not destroy each other, i.e., whether there is a long path (current long path) over this standard checking tree.

If a unit in a 3-unit layer destroys both units in a 2-unit layer, this unit must not have any useful 2SAT paths and thus we temporarily delete it.

Note that a unit's all possible useful 2SAT paths may be exponential, but its all possible useful units are polynomial ($O(n)$ literals). We do not have to remember its all possible useful 2SAT paths but only remember its all useful units and all useful unit pairs. To calculate a 2SAT path, the time is $O(m)$. For each unit in 3-unit layers, we may calculate more than one 2SAT paths which contain different units and different unit pairs. But in average, the time to calculate each unit's useful units and useful unit pairs is $O(n)$* $O(m)$.

In the same way, we also calculate whether *x* destroys *y* when we add the unit $a_{i+12}$, $a_{i+13}$. Only when in each of the three cases, *x* always destroys *y*, then *x* destroys *y* finally.

Please note: in this algorithm and proof, the indirect destroying is based on the old checking tree, i.e., before the unit $a_{i+11}$ is added.

We call a long path before the trouble units being deleted an **old long path** and call a long path after the trouble units being deleted a **real long path**.

Next, we discuss how to calculate a standard checking tree to get a long path if it exists.

### 2.2 Calculate a standard checking tree

**A part path over some layers** means that each unit in the part path is exactly in each of these layers and these units do not have contradictions.

We call each layer in 2-unit layers **an ancestor layer** to all units in 3-unit layers. A layer in a higher layer in 2-unit layers is an ancestor layer to all units in lower 2-unit layers. A 2-unit layer is also called a parallel pair.

If each unit in a part path in the 3-unit layers has the same one 2SAT path, we say that **this part path has this 2SAT path**.

For the standard checking tree, we have 2-unit layers and 3-unit layers. If the 2-unit layers contain only one 2SAT path, when each unit in the rest three-unit layers has this useful 2SAT path, thus, there is a real long path. If they contain two 2SAT paths, we call these two 2SAT paths **two parallel 2SAT paths.** The two paths are utterly different or may share a single part path. Even if they are different, some literals in them may be the same. This does not affect our method. If they contain more than two 2SAT paths, we first calculate the parallel 2SAT paths.

For convenience, suppose the 2-unit layers do not contain 1-unit layers, if they contain, because none of other units destroys any one unit in the one-unit layers, this does not affect our method and proof.

If a unit destroys any one unit in a part path, we say that **the unit destroys this part path**. Also if a unit in a part path destroys a unit in another part path, we say that the former part path destroys the latter part path.

**Algorithm E**: calculate a standard checking tree and try to get a long path

Input: a standard checking tree, in the process of calculation, we call it the **calculating main checking tree.**

Output: a long path over the checking tree, or an answer: no long paths.

Data structures: remember the current calculating main checking tree, including 2-unit layers and 3-unit layers, all contradiction pairs and additional contradiction pairs. For the calculating main checking tree, remember each final leaf with its destroying parallel pairs, and its good units, a calculated unit's redundant units, and the new produced additional contradiction pairs for a calculated unit, that is, the additional contradiction pairs caused due to deleting its redundant units.

1) At first, there are some 2-unit layers. They are **the first generation** 2-unit layers and units in these layers are **the** first generation units. Their sons are second generation, and so on. The first generation is the **highest** generation. The first layer is the **highest layer**.

2) Then choose one unit $u$ from the lowest layer in the first generation, find all units in 3-unit layers each of which destroys (all directly and all indirectly destroys, including additional contradictions) the unit $u$. We call them ***u*'s redundant units** or the unit $u$ **has these redundant units** and call $u$ the current calculating unit. Delete the redundant units and move the rest units in these layers to 2-unit layers and we call these layers **new 2-unit layers for *u***. Then call algorithm 1 to get a standard checking tree. Then we call $u$ **a calculated unit**. For a unit, if we have checked the standard checking property for it (no matter the property holds or does not hold) after its redundant units were deleted, we call it **a calculated unit**. In the mean time, some 3-unit layer may become 2-unit layers. They are also new 2-unit layers and the deleted units in these layers are also redundant units. In the 2-unit layers, we only consider such layers which contains two units. If a layer only contains one unit, then when the standard checking tree property holds, this one unit does not destroy any other units. So this does not affect our method. We call this **the one unit layer property**. The 2-unit layers include 1-unit layers.

3) After the standard checking tree property holds, we call units in the new 2-unit layers the unit $u$'s **son (or descendant) units** and call the new 2-unit layers $u$'s **son 2-unit layers**. Choose one son unit to calculate. We firstly choose such a son $x$ to calculate whose layer's third deleted unit is exactly destroyed by $u$ (that is, this third unit is destroyed by $u$ just before the third unit in $u$'s layer was deleted). And also we firstly try to choose such a son $x$: for any one unit in the old 2-unit layers whose redundant units are also $u$'s redundant units, we can find a unit in the son 2-unit layers whose layer's third unit is exactly destroyed by that unit and whose redundant units are also $x$'s redundant units. Let $y$ be the other unit in the same 2-unit layer with $x$. We find the longest part path in the new 2-unit layers which contains $x$ and in which every other unit's destroying units (these units are exactly destroyed by this unit) in 3-unit layers are $x$'s redundant units. We call this part path $x$'s single part path. We also get $y$'s single part path. Now whose single part path is shorter (contains less units), we choose whom as the son to calculate. We call this **the shorter single part path rule**. If both are the same long, choose any one. Then get the son's son to calculate. Recursively do so.

4) When the standard checking tree property holds and before we calculate the first unit in the first generation, we call all contradiction pairs (contradiction units) including indirect and additional contradiction pairs at this time **the original contradiction pairs**. For an original contradiction pair, one **originally destroys** the other one. Redundant units based on original contradiction pairs are **original redundant units**. For a unit *o*, consider such layers which only contain *o*'s original redundant units (the third units are redundant units), units in the first two units in such layers are *o*'s original son units.

5) A part path over the whole first generation 2-unit layers is a useful 2SAT path and we call all unit in this path a **whole set of units**. For the units in a part path over the new 2-unit layers after deleting a unit *u*'s redundant units, we call them a **whole son set of units of *u***. In this way, a son also has sons. We call any two units in the same whole set **brother units**. For each unit in a real long path, we call it a **correct unit**. Otherwise, it is a **wrong unit**. At first, we do not know which is correct or wrong. If two units have the same father and may have contradictions, we call them **brothers in law**.

6) After we calculated a unit, deleted its redundant units and then the standard checking tree property holds, we take its one son to calculate. If the son does not have redundant units, or, it has redundant units but after all its redundant units were deleted and the standard checking tree property holds and then there are no 3-unit layers, or, after all its redundant units were deleted and the standard checking tree property holds and then in its son units, there is a whole son set in which each unit is a finished unit, then it is a **finished unit.** We will see: a correct unit at last always can be a finished unit when its ancestors are correct. **After some redundant units are deleted, correct units (i.e., real long paths) may become less, but this does not affect our method, because we only concern the last real long paths. The last real long paths may be more than one, but this still does not affect our method.** Considering the **first calculated wrong unit** (that is, before it, all calculated units are correct), we call the new 2-unit layers after its redundant units were deleted **the first occurred wrong 2-unit layers**, that is, before the redundant units in these layers being deleted there are long paths over all layers and after they were deleted, there is no long path. These long paths are **the last real long paths.** We will see that each time we only concern one last real long path and only take units in this long path as **correct units** and all other units are wrong units. Also we call the first 2-unit layer which contains two wrong units **the first occurred wrong 2-unit layer**, that is, each 2-unit layer higher than this layer contains at least one correct unit.

7) If after we delete the current calculating unit's redundant units and move the rest units of these layers to 2-unit layers, the standard checking tree property cannot hold, including that in some layers all three units are deleted in the process of calculating the standard checking tree, then the unit is not a finished unit. So that the standard checking tree property can hold is the key. We only delete the redundant units in 3-unit layers but not in 2-unit layers (note that if a unit in 2-unit layers is not in a useful 2SAT path, it has to be deleted), because the standard checking tree property can guarantee the 2-unit layers. When the standard checking tree property holds and a unit only destroys some units in 2-unit layers and does not destroy any units in 3-unit layers, then this unit is a finished unit.

8) If in a whole set of units in the first generation, each unit is a finished unit, there is a real long path.

9) For a unit, if after the redundant units for it are deleted there are no 3-unit layers or if it does not have redundant units and there are no 3-unit layers, then when the standard checking tree property holds, there is a real long path. If it does not have redundant units and there are 3-unit layers, we have to calculate its unfinished brother units. If no such brother units (its ancestors also do not have unfinished brother units), there is a real long path.

10) If a unit at last cannot be a finished one, we recover all deleted units for it including set the new additional contradiction pairs after calculating it to no contradiction and then calculate its parallel pair mate. If both are not finished, then their father is not a finished unit and thus we calculate the father's parallel pair mate if it was not calculated (in fact, if the standard checking tree property does not hold after deleting the redundant units of each of the two units in the same 2-unit layer respectively, we do not calculate the father's parallel pair mate, but do the algorithm E1. This can increase the calculation efficiency). If a unit cannot get to be a finished one, it is not certainly a wrong unit unless it is in the first generation.

11) After we have calculated one or more units which are finished units, **how to choose a brother unit to calculate?** The finished brother units and their finished descendants shape a part path. For other 2-unit layers in each of which only one unit does not destroy this part path, do not have to calculate such layers. For other 2-unit layers in each of which both two units do not destroy this part path, choose any one brother unit to calculate. If this one cannot get to be a finished unit, then calculate its parallel pair mate. If both cannot get to be finished units, then the father unit is not a finished one. If a parallel pair in the first generation are not finished units, there is no real long path. If after we got a finished unit and the standard checking tree property holds, there are only 1-unit layers and 3-unit layers, there is a real long path and we choose any one 3-unit layer, delete any two units in it and calculate the rest one unit.

12) In summary: At first, for each useful 2SAT path, all units it contains are a whole set of units. They are brother units (may change later). Take one unit, say, $s_1$, and find all units in the 3-unit layers which destroy $s_1$. We call these units $s_1$'s redundant units and call the redundant units' layers $s_1$'s **son layers** or **new 2-unit layers**. Move the son layers to the 2-unit layers and delete the redundant units. Then do algorithm 1 to let the standard checking tree property hold. If it cannot hold, $s_1$ is a wrong unit. In the mean time, some units may be deleted, but based on the standard checking tree property, this does not affect our method. Then we call all units in the new 2-unit layers $s_1$'s son units. For a part path over the son layers, we call all units it contains brother units. Take a son unit, say $s_2$, to calculate. **Note that** we firstly choose such a son whose layer's third deleted unit is exactly destroyed by $s_1$ (that is, this third unit is destroyed by $s_1$ just before the third unit in $s_1$'s layer was deleted). Find the units in 3-unit layers which destroy $s_2$. We call them $s_2$'s redundant units. These layers are $s_2$'s son layers. Move these layers to the 2-unit layers and delete the redundant units. Let the standard checking tree property hold. If $s_2$ together with $s_1$'s deleted redundant units in other layers shape a part path, that is, these units do not destroy each other, this is a problem, because $s_2$ together with $s_1$'s such deleted redundant units may be in a real long path. We call it t**he separated whole redundant units case** or t**he separated whole redundant units problem**. We call $s_1$'s such redundant units the **separated redundant units** and call $s_2$ the **separated redundant units' brother**. We will solve this problem later. Except for the separated whole redundant units case, a unit's redundant units are as a whole, that is, this case does not exist: part of the redundant units are in a real long path and the rest part are not in this real long path. In this way, recursively, we can get $s_2$'s son, son's son, and so on. If we get that a unit is a finished unit, **there are three cases for a finished unit**. Case 1: there are 2-unit layers and 3-unit layers, and then we take one unfinished brother unit of the finished unit to calculate as stated above. Case 2: if there are 2-unit layers and there is at least one part path over the 2-unit layers which only contain finished units, then only keep this one part path in 2-unit layers. So at last there are only 1-unit layers and 3-unit layers, and at this time, we choose any one 3-unit layer, delete any two units in it and calculate the rest one unit. Case 3: there are no 3-unit layers, and this time we can directly get a real long path. If a unit failed, that is, it cannot be a finished unit, we recover all units deleted for it and also recover new additional contradiction pairs to no contradiction and then calculate other units.

13) For the current calculating unit $x$, if it is a correct unit and it cannot get to be a finished one, then we can determine that it is destroyed by one or more parallel unit pairs (destroying parallel pairs including hidden destroying parallel pairs which we will explain later to make some correct units be additional contradiction pairs) which are brothers (or brothers in law) of $x$'s some ancestors in 2-unit layers, that is, a part of the real long path which contains $x$ is destroyed by one of the parallel pair and another one part of the real long path is destroyed by the other one of the parallel pair. We call such parallel unit pair a destroying parallel unit pair or **a destroying parallel pair** to $x$ or it is $x$'s a destroying parallel pair and $x$ has this destroying parallel pair. Destroying parallel pairs **do not include** the first generation units.

14) Except original contradiction units, other contradiction units are additional contradiction units. We calculate and decide new additional contradiction units after deleting a unit's all redundant units. Based on this, we can get the destroying parallel pairs for each later calculating unit. If after a unit's redundant units are deleted, the standard checking tree property does not hold, we call it **a basic unfinished unit**. It is a leaf. If it does not hold only due to original contradiction units, we call it **a leaf of kind 3.** If it does not hold due to additional contradiction units after some redundant units were deleted, it is **a leaf of kind 1**. If a unit's all descendants have been calculated at other place, it does not have more descendants here and thus we do not need to calculate its descendants again, we call it a **leaf of kind 2**.

The leaf of kind 3 is a wrong unit. In algorithm E1, we change the concept "leaf", and we define the leaf calculated in algorithm E1 **a final leaf**. Later our all recalculations are based on the final leaves.

15) For not-calculated units, if a unit has two parallel sons (or only one son), we choose one son to calculate. If it cannot get to be a finished one, we calculate the other one in the same layer if it exists and was not calculated. Note that for these two units in the same layer, **if possible we always choose such a unit to calculate when deleting whose redundant units the standard checking tree property holds**. If both cannot get to be finished, we calculate the unit in the same layer with the father if it exists and was not calculated (in fact, for efficiency, we do algorithm E1). We call all these **the unit calculation order**. Recalculating units is also according to the calculation order. Note that only when a unit and the other one unit in the same layer both are leaves (that is, when delete such a unit's redundant units the standard checking tree property does not hold), we take each of them as a leaf to calculate (do the algorithm E1 to calculate the final leaf and its good units, about good units, see algorithm E1). We call this the success unit rule.

16) We calculate a unit, delete its redundant units and then the standard checking tree property does not hold. Then we do the algorithm E1 to get the final leaf with its good units.

17) The recalculating units rule and the recalculating property. After did algorithm E1, we got a final leaf with its good units (if algorithm E1 failed, that is, no such leaves and goal generations, then algorithm E failed, that is, no long path for this time, then we calculate the next unit). It is **the final leaf**. If the final leaf's goal generation is the generation which contains the first occurred wrong 2-unit layer, then we can get that the final leaf's good units are correct units, and then, in all later calculation, these good units are always in 1-unit layers and do not change (because we always try to calculate units in lower layers). We call such a leaf **a successful final leaf**. In the final leaf's good units' layers, we delete the first two units and only keep these good units. Note that at this time, we define that **the exact contradiction pairs** are the contradiction pairs just before the third unit in the final leaf's goal generation's father's layer was deleted. Then we calculate the final leaf's goal generation's father in law as a unit, delete its redundant units, let the standard checking tree property hold and then calculate its son, and son's son according to the calculation order. Until get finished units or until when we calculate a unit and delete its redundant units the standard checking tree property does not hold. Note the success unit rule. Then we do algorithm E1 to get a new final leaf. This is also **the precondition for recalculating**, that is, under this one final leaf, we can recalculate or calculate any one unit one time. Continue in this way until get finished units. In this way, for each final leaf, we calculate a unit at most one time. Note that if the final leaf's goal generation is lower than the first occurred wrong 2-unit layers, the final leaf with its good units may be wrong units. If so, the deleted units in the good units' layers must be recovered later in algorithm E1. Then we call this leaf **a failed final leaf**. Consider the goal generation discarding rule (see algorithm E1) and that for each leaf with the same good units, we recalculate a unit at most one time, failed final leaves do not affect the polynomial. A failed final leaf (if it is a correct unit, but its goal generation is lower than the first occurred wrong 2-unit layers) may become a successful final leaf when the goal generation becomes higher and the good units become more. Then we can calculate units for this new leaf again.

18) In this way, if there are real long paths, we always can get one. If after the algorithm E1, there is a part path over all 2-unit layers and all units in it are finished units, and also there are still some 3-unit layers, then we delete any two units in any one 3-unit layer and then calculate the rest one unit.

19) **The thinking thread of the algorithm E**: do this algorithm according to the unit calculation order. When we calculate a unit, delete its redundant units, if the standard checking tree property does not hold, we calculate the other one unit in the same 2-unit layer with this unit. If for both units, the standard checking tree property does not hold, we do the algorithm E1 to calculate the final leaf. If E1 fails, there is no long path for this time. If E1 succeeds, we get the final leaf. In this leaf's good units layers, only keep the good units. Put these layers into 2-unit layers at the bottom. Then we calculate the final leaf's goal generation's father in law as a unit, delete its redundant units, let the standard checking tree property hold and then calculate its son, and son's son according to the calculation order, and so on. Until get finished units or until do algorithm E1 to get a new final leaf.

20) When without hidden destroying parallel pairs, if after we deleted a unit's redundant units the standard checking tree property does not hold, we call the new 2-unit layers (the redundant units' layers, before these redundant units were deleted, they were destroyed by the unit) **a last generation**. But if there are some hidden destroying parallel pairs, when we get such a last generation, the standard checking tree property may still hold. This time we call this last generation **a special last generation**. We discuss the special last generation and hidden destroying parallel pairs in algorithm E1. When we get a special last generation, we still do algorithm E1 to calculate the final leaf.

PROOF.

Note that in the algorithm 1, if a unit's redundant units are deleted in 3-unit layers (we call these layers new 2-unit layers), we move these layers to 2-unit layers, and in the process of calculation, we have to move such 3-unit layers in which some units do not have useful units (delete such units) to 2-unit layers and then recalculate.

We first see the following lemmas:

LEMMA 1: *from a destroyed checking tree to a standard checking tree, we may have some new additional contradiction pairs, and we also may delete some units, but this job does not lose any real long paths.*

From the calculating process (the following algorithm 1) we can see: for any one real long path, any two units in it cannot become contradiction unit pair and any one unit in it cannot be deleted.

LEMMA 2: *after we have got one or more finished units (they are brothers) in the first generation, these units and their descendant finished units shape a part path. Units in the rest 3-unit layers do not destroy this part path. For each of the other 2-unit layers, if one unit destroys this part path, then the other one does not destroy this part path and also does not destroy any units in the rest 3-unit layers. All such one unit (not-destroying unit) together with this part path shape a longer part path. Any units in this longer part path do not destroy any units in the rest 2-unit layers and 3-unit layers. The rest 2-unit layers are the 2-unit layers before the calculation, that is, before any redundant units being deleted. So if before the calculation, there are real long paths, now there are still real long paths which contain the longer part path. We call this **the holding real long path property**.*

From the process of the calculation, we can see that the lemma 2 holds. Note these facts: let $d_1$, $d_2$ be two parallel units (a parallel pair) in 2-unit layers. Let $p_1$, $p_2$ be two part paths in 3-unit layers. When the standard checking tree property holds, if $p_1$ destroys $d_2$ but does not destroy $d_1$ and $p_2$ destroys $d_1$ but does not destroy $d_2$, then $p_1$ destroys $p_2$. If $p_1$ does not destroy $p_2$ and they are in one part path, and $p_1$ destroys $d_2$ but does not destroy $d_1$, then $p_2$ also does not destroy $d_1$. Let $d_3$, $d_4$ be two parallel pair units. $d_3$ destroys $d_2$ and does not destroy $d_1$. If $p_1$ and $p_2$ do not destroy $d_3$, then $p_1$ and $p_2$ also do not destroy $d_1$.

LEMMA 3: *at any time, when the standard checking tree property holds, if there is a part path over all the 3-unit layers, then there is a real long path.*

This is because that if units of this part path do not share the same one useful 2SAT path, some units of them must have additional contradiction and thus cannot shape a part path.

Based on the lemma 2 and lemma 3, we can get the lemma 4 and lemma 5.

LEMMA 4: *if we always can keep the rest part of the part path in lemma 3 as a part path over the updated 3-unit layers, at last, we can get a real long path. Thus, the first occurred wrong 2-unit layer(see the above 6 in algorithm E) is the key.*

LEMMA 5: *if before the calculation, that is, just at the beginning of the algorithm E, there were real long paths, and now when some redundant units have been deleted there is no part path over the 3-unit layers, and a correct unit cannot get to be a finished unit, then there are some destroying parallel unit pairs in higher 2-unit layers which let some units in the last real long path become additional contradiction unit pairs (especially let the part path over 3-unit layers* *become a non-part path, that is, some units in it become to have contradictions), that is, the correct unit is destroyed by some brothers or brothers in law of its some ancestors, so that it cannot get to be a finished one.*

Note the lemma 5, if a correct unit cannot get to be a finished one, this means it is destroyed by the brothers or brothers in law of its some ancestors, that is, some higher 2-unit layers (destroying parallel pairs including hidden destroying parallel pairs to make some correct units be additional contradiction pairs.). Only for the separated whole redundant units case, a correct unit may be destroyed by its own brothers or brothers in law (as destroying parallel pairs), but this does not affect our method. A wrong unit may be destroyed by correct units, but when it fails to be a finished one, we recover the deleted units for it. Note that a correct unit may have both correct and wrong descendants, but this does not affect the above rule.

LEMMA 6: *in this calculation, layers above the first occurred wrong 2-unit layer are not changed and thus the first occurred wrong 2-unit layer only can become lower and lower*.

Considering the **first calculated wrong unit** (that is, before it, all calculated units are correct), we call the new 2-unit layers after its redundant units were deleted the first occurred wrong 2-unit layers, that is, before the redundant units in these layers being deleted there is a part path over the 3-unitl layers and after they were deleted and when the standard checking tree property holds, there is no such a part path due to the additional contradiction pairs. Because we always concern destroying parallel pairs in the lowest layers, this can make the first occurred wrong 2-unit layer become lower and lower, that is, after we change the first calculated wrong unit to a correct unit to calculate, it will not be changed back to a wrong one again. Though a wrong unit's destroying parallel pairs may be in higher layers than the first occurred wrong 2-unit layer, before return to these higher layers, we certainly can get destroying parallel pairs to destroy some correct units which are in the layers not higher than the first occurred wrong 2-unit layer. **A correct unit** means that **just** before the redundant units of the first calculated wrong unit were deleted, it was in a real long path. Also note that: the first occurred wrong 2-unit layer is as a destroying parallel pair to destroy some correct units, because if they are not, there is currently a real long path. We only concern the one last real long path in the above 6) of the algorithm E and units in such a real long path are correct units. Other units are **wrong units**. Correspondingly, we concern **the last part path** over the 3-unit layers which are contained in the last real long path. For the first time we calculate a wrong unit, we call this wrong unit **the first time's first calculated wrong unit**. Also we have **the first time's first occurred wrong 2-unit layers**.

For *k* units in *k* layers, if some of them are contradiction units, we call them **a false part path over these layers**

We call all contradiction pairs at the beginning of the algorithm E **original contradiction pair**s. Later new additional contradiction pairs are not original, but are **caused contradiction pair**s caused by ancestor destroying parallel pairs.

For two original contradiction units, we say one **originally destroys** the other one. For two additional contradiction units, we say one **additionally destroys** the other one.

If we calculate a unit, delete its redundant units and check whether the standard checking tree property holds, we call it **a calculated unit** and call a layer which contains calculated units **a calculated layer**.

In a 2-unit layer, we call the unit which was in this layer but was deleted as a redundant unit **the third deleted unit** or **the third redundant unit**.

LEMMA 7: *The precondition for recalculating units (in algorithm E) for a final leaf. For each final leaf, we recalculate a unit at most one time. So a unit is recalculated O(n) times. The recalculating job is polynomial.*

When we calculate according to the calculation order, **the precondition for recalculating units for a leaf is**: after the algorithm E1 succeeds, we get a final leaf, delete the first two units in its good units' layers, delete its redundant units, and the standard checking tree property holds. Then we can calculate or recalculate each unit at most one time for this leaf according to the calculation order in algorithm E.

Note that only when the leaf *l* and the other one unit in the same layer both are leaves, we take one of them as a leaf to calculate (do the algorithm E1 to calculate the final leaf and its good units).

We can see: there is at least one correct final leaf, its good units are correct units and its destroying parallel pairs are wrong units and we can get all these. When we recalculate a correct unit, if the correct unit cannot get to be finished this time, we can get a new correct leaf of kind 1 and we can recalculate for this leaf. In this way, correct units always can be combined with correct units at last.

When we recalculate a calculated unit, it may get to be a finished unit , or it has new leaves of kind 1, or we can determine it is a wrong unit and then do not calculate it later.

If a part path is not utterly in any one real long path, then when the standard checking tree property holds, there are three or more long paths (old long paths), and any two units in the part path are in at least one of these long paths. We call this **the three or more long paths property**. Note this special case: when we calculate a unit, its third redundant units may contain only correct units and the first two units in these layers may only contain wrong units, and vice versa.

By the following five points, we can know that the recalculating job is polynomial.

1) If a final leaf's goal generation contains the first occurred wrong 2-unit layer, then we always can get that the final leaf's good units are correct units, its good units will always be in 1-unit layers and will not change after the leaf was calculated in algorithm E1. If the final leaf's goal generation is lower than the first occurred wrong 2-unit layers, and if its good units contain wrong units, we will have the opportunity to recover correct units in these layers in algorithm E1. Consider the goal generation discarding rule and that for each final leaf we recalculate a unit at most one time in algorithm E. This is the key and it makes things easy.

2) At anytime, we do not need to consider a lot of (exponential) combinations of different good units.

3) If a correct calculated unit has been recalculated for a final leaf *cl* under a wrong unit, at last we always can get a new final leaf which is caused by the wrong unit (it destroys some third correct units). Then we have the opportunity to recalculate this correct calculated unit for a correct leaf. Then by the recalculating property, we do not have to repeat to recalculate this unit for *cl*.

4) A unit is never recalculated for the same leaf with the same good units more than one time, because if the unit is a correct unit, in the first recalculation, this unit can get more descendants and new correct leaves or is finished.

5) A key problem is: **for different final leaves with their good units, how to get a combination of only correct final leaves whose good units are correct units in polynomial time**? By the first correct leaf property, the one to one final leaf combination rule, the correct chosen units property and the final leaf combination property (see algorithm E1), we can get this combination in polynomial time.

How to get the 2SAT's polynomial and to avoid the 3SAT's exponential?

1) The two parts: the destroying parallel pairs and the third units, and the partial 2SAT logic.

2) For some 3-unit layers in which the third units are wrong units, we do not need to choose the first two units two times (each time choose one) and thus make 3***m* combinations, but only delete the third units (redundant units) and then calculate the first two units based on the standard checking tree property.

3) For layers whose third deleted units are correct units, we have the opportunity to get a correct leaf with its correct good units.

**4)** We do not recalculate a unit for the same leaf more than one time in algorithm E.

When the first occurred wrong 2-unit layer's third correct unit is recovered and calculated, it will not be changed, that is, it will always be an ancestor of all later calculated units. By the algorithm E1, we always have the opportunity to let the first occurred wrong 2-unit layer be as a destroying parallel pair and the third correct unit in this layer be a good unit for a final leaf and the final leaf's goal generation is the generation which contains the first occurred wrong 2-unit layer.

**Algorithm E1**: calculate the final leaf and its good units

Input: the calculating main checking tree with a leaf of kind 1 *l*, *l*'s redundant units

Output: the final leaf and its good units, the first two units in the leaf's good units' layers are this leaf's destroying parallel pairs. Only keep the final leaf's good units in their layers, when the standard checking tree property holds, keep this time's calculating main checking tree, then with this tree return to algorithm E, also keep the data of other last chosen units with their ancestor chosen units which have the current goal generation

For the current calculating unit *l,* if it is a correct unit and after its redundant units are deleted the standard checking tree property does not hold, some correct units in 3-unit layers must be destroyed by one or more destroying parallel pairs in 2-unit layers, that is, a part of the last real long path is destroyed by one of the parallel pair and another one part of the last real long path is destroyed by the other one of the parallel pair. Suppose *a*, *b* are two units in a 2-unit layer and *e*, *f* are two correct units in different 3-unit layers. *a* destroys *e* and *b* destroys *f*. The destroying parallel pair *a, b* makes some correct units such as *e*, *f* destroy each other. *c*, *d* are two units in a higher 2-unit layer and *c* destroys *b* and *d* destroys *f*, that is, *b*, *f* are an additional contradiction pair caused by *c*, *d*. We say *c, d* are *a, b*'s one **causing destroying parallel pair** or one **causing destroying layer** and *a, b* are *c, d*'s one **caused destroying parallel pair** or **one caused destroying layer**. A caused destroying parallel pair may also be a causing one and vice versa. We do not know which one is a correct or a wrong unit.

We have direct contradiction pair, indirect contradiction pair and additional contradiction pair. Now we define a new concept "**exact contradiction pair**" or "exact destroying". We take an example to explain the concept exact contradiction pair. Let *a*, *b*, *c* be three units in a 3-unit layer. So are units *d*, *e*, *f*. We take a unit *u* in the 2-unit layers to calculate. Units *c* and *f* are *u*'s redundant units. After the redundant units are deleted, the standard checking tree property holds and the layer of *a*, *b* and the layer of *d*, *e* are the new 2-unit layers. Let *x*, *y* be two units in different 3-unit layers which were not a contradiction pair before the redundant units *c* and *f* were deleted but are a contradiction pair after they were deleted. So *x* and *y* are an additional contradiction pair. If after we only delete *c*, *x* and *y* are an additional contradiction pair, then *a*, *b* are a destroying parallel pair to destroy *x* and *y* (to exactly destroy, that is, *a* exactly destroys *x* and *b* exactly destroys *y*). If only after we delete both *c* and *f*, *x* and *y* are an additional contradiction pair, then both *a*, *b* and *d*, *e* together are (exact) destroying parallel pairs to destroy *x* and *y*. If after we only delete *f*, *x* and *y* are not an additional contradiction pair and after we delete both *c* and *f*, *x* and *y* are an additional contradiction pair, then *a*, *b* are a (exact) destroying parallel pair to destroy *x* and *y* but *d*, *e* are not. At this case, suppose that *a* destroys *d* and *b* destroys *e*. Though after *c* and *f* are deleted, *a* destroys *x* and *b* destroys *y,* and also *e* destroys *x* and *d* destroys *y*, but we say that *a* exactly destroys *x* and *b* exactly destroys *y,* and do not think that *d* exactly destroys *y* and *e* exactly destroys *x.*

After we deleted redundant units and got new 2-unit layers, how to decide each exact contradiction pair between the new 2-unit layers and 3-unit layers? Any two units between the new 2-unit layers and 3-unit layers (one unit is in the new 2-unit layers and the other one is in 3-unit layers) which was a contradiction pair (direct, indirect or additional) just before the redundant units were deleted is an exact contradiction pair.

Note this case: units $a_1$, $a_2$ are in a 2-unit layer and the third deleted unit is $a_3$. So are $b_1$, $b_2$ and the third unit is $b_3$. $c_1$, $c_2$ are two correct units in different 3-unit layers. Without $a_3$, $b_1$ and $b_2$ destroy $c_1$ and $c_2$ respectively and without $b_3$, $a_1$ and $a_2$ destroy $c_1$ and $c_2$ respectively. That is, both $a_1$, $a_2$ and $b_1$, $b_2$ may contain correct units when $a_3$, $b_3$ exist. This time, for the same logic as the last real long paths, we choose the higher layer's first two units to contain correct units and the lower layer to contain only wrong units. We call this **the lower wrong 2-unit layer rule**.

Now the standard checking tree property holds and the last real long path (can be more than one, but we only concern one) exists. Then we take a wrong unit *u* to calculate. We delete *u*'s redundant units and the standard checking tree property holds. But now there is no real long path over all layers. We call the new 2-unit layers (that is, the layers which contained *u*'s redundant units formerly) the first occurred wrong 2-unit layers as stated formerly and call this generation **the current first generation of wrong 2-unit layers**. We do not know which generation is the current first generation, but it exists.

A unit's redundant units may be destroyed by different destroying parallel pairs (by one unit of each pair). A correct unit's redundant units may contain correct units, but these correct units at last are destroyed by destroying parallel pairs which are wrong units not by the unit itself.

Suppose that we calculate a unit *u* in algorithm E. It is possible that there are some destroying parallel pairs which contain only wrong units and which as destroying parallel pairs destroy a part path of correct units into two part paths in 3-unit layers but which do not destroy *u*. We call such destroying parallel pairs **hidden destroying parallel pairs**. The destroying parallel pairs which cause *u* and units in 3-unit layers to be additional contradiction units and/or one unit of this pair destroys *u* are **visible destroying parallel pairs**. We do not know whether a parallel pair only contains wrong units, but this does not matter and they are still hidden destroying parallel pairs for the above case, that is, for two units in a 2-unit layer, after we calculate the current calculating unit *u* and delete its redundant units, if every one of the two units still destroys some units in 3-unit layers, they are a hidden destroying parallel pair. A hidden destroying parallel pair may become a visible destroying parallel pair later.

Note that if without hidden destroying parallel pairs and there is only one real long path, exact contradiction pairs are the contradiction pairs just after we got the first occurred wrong 2-unit layers. The difference between two definitions does not affect our method. Next when we calculate a chosen layer and its father layers, we only consider exact destroying.

In a destroying parallel pair in 2-unit layers, if only one unit does not destroy *u*, then units destroyed by this unit are *u*'s redundant units. Note that after *u*'s redundant units are deleted and thus there are some new destroying parallel pairs, we calculate the standard checking tree property. When hidden destroying parallel pairs being considered, all units destroyed by destroying parallel pairs are considered.

Then we take a unit in the new 2-unit layers to calculate. Delete its redundant units and let the standard checking tree property hold. We call this time's new 2-unit layers the current second generation of wrong 2-unit layers. It is a lower generation than the first generation. Then we may have the third generation, fourth generation, and so on. These jobs are in algorithm E. Each time the standard checking tree property holds. Note that each generation contains some destroying parallel pairs which are only wrong units (consider the lower wrong 2-unit layer rule). If it does not contain, the hidden destroying parallel pairs must contain. Also note that a generation may contain one or more finished units and we still call the new 2-unit layers of the last calculating unit which is not a finished one in this generation the next generation wrong new 2-unit layers.

For a new 2-unit layer, if its third deleted unit destroyed by a destroying parallel pair (by one unit of the pair, exact destroying), then we call the layer of the destroying parallel pair **a father or an ancestor layer** of the new 2-unit layer and call the latter a **son or a descendant layer** of the former. Note that a hidden destroying parallel pair's sons may not be in the next generation currently.

Now we calculate a unit *l* (in algorithm E), delete its redundant units and get its new 2-unit layers. The standard checking tree property does not hold. So *l* is a leaf. We call the generation which only contains *l*'s redundant units' layers **the current last generation of wrong 2-unit layers**, or **the last generation**. These redundant units are destroyed by *l* before they are deleted.

The 3-unit layers contain two parts. The first two units are the first part and the third units (which as redundant units) are the second part. The destroying parallel pairs take up the first part. These are the base to constitute a 2SAT. The key problem for our algorithm is how to solve the separated whole redundant units problem. Obviously, if without the separated whole redundant units case, that is, a unit's redundant units are whole correct (in a real long path, and no real long path contains only part of the redundant units) or are whole wrong, we can calculate the units in polynomial time and the calculation of units has the same logic as 2SAT. When this case happens, there is at least one layer in the current first generation of wrong 2-unit layers whose two units are wrong units, this layer also has at least one son layer whose two units are wrong units (considering hidden destroying parallel pairs, also if two layers' first two units contain possible correct units but they are not correct units in the same real long path, we let the lower layer's first two units be wrong units, see the lower wrong 2-unit layer rule), and so does the son, the son's son.... The first occurred wrong 2-unit layer contains two wrong units and later each new first occurred wrong 2-unit layer (after it is updated) contains two wrong units. So partially we still can get 2SAT logic only based on these layers whose first two units are wrong units and the third units are correct units**.** We call this **the partial 2SAT logic.** We call a layer whose first two units are wrong units and whose third unit is a correct unit **a partial 2SAT layer**. When we get the last generation, we concern such 2-unit layers whose two units are wrong units and these two units destroy at least two correct units in 3-unit layers respectively (one unit destroys one).

**Note that the key foundation for the polynomial for the whole algorithm is the partial 2SAT logic**.

Now we have two problems: 1) when the standard checking tree property does not hold, we need to find the layers in which the third deleted units are correct units, but how to separate the partial 2SAT layers from other layers; 2) the causing and caused destroying layer problem. We call them together **the partial 2SAT logic problem**.

**The way to get the partial 2SAT:**

The key for this way is: based on the strong function of the standard checking tree property and the importance of the first occurred wrong 2-unit layer, we can independently calculate each layer respectively and get the first generation of wrong 2-unit layers at last. At first we do not know the current first generation of wrong 2-unit layers.

The method to get the partial 2SAT and to calculate chosen units is complicated and thus now we firstly give a rough thinking thread.

In algorithm E, when the standard checking tree property does not hold, at first, we choose a layer in the current last generation to calculate and we call it the current last chosen layer, or the last chosen layer (now we start the algorithm E1). Then if the standard checking tree property does not hold, calculate its father layers in an upper generation and they may be more than one. We may calculate each father layer respectively, but consider them (the same generation) together. Then for these father layers' father layers, we do them in the same way.

In algorithm E1, when check the standard checking tree property for a last chosen unit and its goal generation (see next), we need to do one more thing and we call it **the additional job for checking**: for two units in different 3-unit layers, at first they have a same one useful pair and they do not destroy each other. If at any one other 3-unit layer, no unit which has this useful pair and also does not destroy these two units, then these two units do not share this useful pair.

Each time, we choose a layer from the current last generation to calculate. It is the current last chosen layer. We delete the first two units and only keep the third unit in this layer. Keep other layers in this generation (keep each one's three units). Then if the standard checking tree property holds (do the additional job for checking in the whole algorithm E1), this generation is **the goal generation**. If the property does not hold, we find the chosen layers' father layers. We also call them the chosen layers in this new generation (one higher than the former generation). Also in this generation keep the chosen layers only with the third units and keep other layers' all three units. For each lower (just done) generation, still keep the chosen layers only with the third units and keep other layers' all three units. Then if the standard checking tree property holds, this generation is the goal generation of the current chosen layers and the highest such chosen layer in this generation is the chosen layers' **destroying layer**.

How to get a chosen layer's father layers? Let the third destroyed unit in this chosen layer be $ud_1$. This layer and other current chosen layers in the same generation have some father layers. One unit in a father layer destroys the third unit in the son layer (they are an exact contradiction pair). Some layers may not have father layers.

About causing and caused destroying layers, when we calculate chosen units as well as their ancestors and find units in the one higher generation which exactly destroy the current chosen units, we only consider this case: two units *a*, *b* are in a 2-unit layer and so are units *c*, *d*. *e* is the third deleted unit in the layer of *a*, *b*. *x* is a unit in 3-unit layers. *d* destroys *a* but does not destroy *b*. Before *e* was deleted, *b* destroyed *x* and *d* did not destroy *x*. After *e* was deleted, *d* also destroys *x* (caused by that *b* destroys *x*). But *b* exactly destroys *x* and *d* does not exactly destroys *x*. So if *x* is a chosen unit, its father chosen unit is *e*, not the third unit in *c*, *d*'s layer. For that case: a destroying parallel pair which are only wrong units cause another destroying parallel pair which contain one correct unit to destroy two correct units which are in 3-unit layers, when calculating chosen units, we consider exact destroying. We do not know whether the destroying parallel pair are wrong units, so no matter whether they are wrong units, we still consider exact destroying (except when the causing destroying parallel pair is in the first generation, that is, original destroying). We call this **the only one time calculation of causing-caused destroying layers property**. This property is very important for the algorithm E1.

So each time for each new 2-unit layer (in the current last generation, each time we choose one layer to calculate and we call it **the current last chosen layer,** or **the last chosen layer**), only recover the third unit in this layer (do not keep other two units in this layer) which was deleted as a redundant unit and also recover the deleted units in old 2-unit layers which were deleted at the same time when the third redundant unit was deleted. We call the third unit in the last chosen layer **the last chosen unit**. For all other layers in this generation, recover the third units, that is, keep all three units in these layers. Check whether the standard checking tree property holds. If the property does not hold, continue to recover this layer's father layers' third units (the father layers in the upper one generation). We call the father layers **the current chosen layers** or the chosen layers in this generation. We call the third unit in each chosen layer **a chosen unit**. It is an ancestor of the last chosen unit. Delete the first two units in these layers and also keep all other layers' three units in this generation and recover the corresponding deleted units in old 2-unit layers. Continue to get to the upper generation in this way until the standard checking tree property holds. Note that when the standard checking tree property holds, some 3-unit layers may become 1-unit layers. We also call this one unit a chosen unit and call these layers chosen layers. All the chosen layers are the ancestors of the last chosen layer and the chosen units are the ancestors of the last chosen unit.

When the standard checking tree property holds, the current generation is **the goal generation**. We call it **the goal generation** of the current last chosen layer (last chosen unit) and of all the current chosen layers (chosen units). Otherwise, we continue to calculate upper generations (until to the first generation). After getting the goal generation, we call the first two units in all chosen layers **the destroying parallel pairs** of the last chosen layer (or last chosen unit) and of all the current chosen layers and call the third units (that is, the chosen units) in all chosen layers **the good units** of the last chosen layer (or last chosen unit) and of all the current chosen layers. We call the highest layer in the chosen layers in the goal generation **the destroying layer** of the last chosen layer (or last chosen unit) and of all the current chosen layers.

After a last chosen layer was calculated, recover the whole tree the same as just before calculating this layer. Then choose another layer which was not calculated in the current last generation to calculate. Until all such layers are calculated.

For a hidden destroying parallel pair whose two units' redundant units in 3-unit layers have not been deleted, when we calculate chosen units in this generation in other layers, the third unit in this layer is recovered in the same way as other layers with no chosen units (in this generation). Also, the third unit in this layer has to be as a last chosen unit to calculate. We call this **the hidden destroying parallel pair property**. The third deleted unit in a hidden destroying parallel pair's layer may be as an ancestor chosen unit.

When we calculate the last chosen unit and its goal generation, we firstly consider the third units exactly destroyed by visible destroying parallel pairs. A third unit may be exactly destroyed by both units of a destroying parallel pair including hidden destroying parallel pair. Then consider the hidden destroying parallel pairs' third units. Also we firstly try to get a final leaf which is destroyed by a visible destroying parallel pair. Note that if a hidden destroying parallel pair becomes a visible destroying parallel pair in a later (lower) generation in algorithm E, then it is not a hidden destroying parallel pair. Take each third unit in a hidden destroying parallel pair as a last chosen unit and calculate its goal generation in the same way. For this kind of last chosen unit, if we get its goal generation, only keep its good units in their layers, and for other 2-unit layers in each generations we do not recover the third units, because the third units cannot be correct units when the goal generation contains the first occurred wrong 2-unit layer (if they are, we can get the final leaf with good units in the third units), but for other hidden destroying parallel pairs which are higher than this last chosen unit, their third units have to be recovered and also those third units which were deleted due to this hidden destroying parallel pair (due to the third unit in this layer being deleted) have to be recovered. At last, we choose the last chosen unit whose goal generation is the lowest as the final leaf. If two last chosen units have the same lowest goal generation, one is a third unit of a hidden destroying parallel pair and the other one is not, choose the one which is not as the final leaf. If two last chosen units have the same lowest goal generation, each one is a third unit of a hidden destroying parallel pair, choose the one which is lower as the final leaf. Note that if the final leaf is a failed final leaf later, we have to try each of other last chosen units as the final leaf. We try each of them also according to the above order for last chosen units which have their goal generations. The new goal generation is a higher generation than the failed generation. We will explain the failed final leaf later. For a hidden destroying parallel pair, if the hidden destroying parallel pair does not become a visible one, we do not need to consider this when calculating chosen units. If the hidden destroying parallel pair becomes a visible destroying parallel pair, treat it in the same way as treating visible destroying parallel pair. Because of this, when we calculate ancestor chosen units, we do not have to consider a lot of causing-caused destroying units. As stated above, we call this **the only one time calculation of causing-caused destroying layers property**. If a hidden destroying parallel pair is higher than the first occurred wrong 2-unit layer, we certainly have other correct last chosen units and if a hidden destroying parallel pair is lower than (not higher than) the first occurred wrong 2-unit layer, we can take the third unit in the pair's layer as a last chosen unit. Also note that after we calculated a unit *u* and then calculate its one son in algorithm E, we firstly choose such a son whose layer's third deleted unit is exactly destroyed by *u*. So, we always can get a correct last chosen unit with its ancestor chosen units which are all correct units.

We calculate every layer's third unit in the current last generation, including layers which contain hidden destroying parallel pairs. For a layer which contains a hidden destroying parallel pair, if no goal generation under its generation can be got, this layer also has to be calculated as a last chosen layer. Until at least one last chosen layer successfully gets its goal generation (other last chosen layers also have to be calculated). At last, if no last chosen unit can get the goal generation, we say that the algorithm E1 fails.

Now we explain the method to get the partial 2SAT and to calculate chosen units in detail.

Our goal is to get the goal generation which contains the first occurred wrong 2-unit layer, but we do not know. So we suppose a goal generation and we call it **the supposed current goal generation** or **the current goal generation**. At first, the last generation's father generation (just above the last generation) is the current goal generation. If no last chosen unit has this goal generation, change it to a higher generation. Later if a last chosen unit's goal generation is lower than the current goal generation, continue to calculate the upper generations until its goal generation is the current goal generation or it does not have this goal generation. Later if all the final leaves which have this current goal generation failed, change the current goal generation to a higher generation. Only when a hidden destroying parallel pair's layer is not higher than the supposed current goal generation, we take the third unit in this layer as a last chosen unit.

In algorithm E, we calculate a unit, delete its redundant units and the standard checking tree property holds. Then we calculate its one son in next generation. Then calculate son's son. Until we calculate a unit *l*, delete its redundant units and the standard checking tree property does not hold. We call the generation which only contains *l*'s redundant units' layers the last generation. This is **the first kind last generation**. Suppose that we always calculate correct units from the start and also suppose that each time we delete the redundant units in the 3-unit layers which are all the destroyed units destroyed by a part path over the whole generation and this part path only contains correct units. If there is such a part path over one generation so that when we delete the units destroyed by this part path in the rest 3-unit layers (if the destroyed units are zero, we consider this generation's father generation) and thus get a new generation, there is a part path over the new generation and this part path is surely in a real long path, then we call the former generation **the second kind last generation**, call the new generation **the last 3-unit layers** and call the two part path together **a key part path**. For a part path over a whole generation in algorithm E, if whether **the three or more long paths property does not fit this part path** (that is, this part path cannot be cut into three or more part paths any two of which is in one different (old) long path), or this part path does not have redundant units in the rest 3-unit layers, or, this part path only destroy some units in upper generations and we can separate the third units which are not all destroyed by this part path (part of them may be destroyed by this part path) from other two units by the third units' speciality in this generation or we cannot separate the third units from other two units in this generation, then we say that this part path **is surely in a real long path**. When we calculate chosen units, we always try to get the key part path unless it has been got. By the calculating process we can see: a key part path is in a long path and if two key part paths do not destroy each other, they are in the same one long path. We call this **the key part path property.**

The current goal generation are the new 2-unit layers whose third deleted units are a unit $u_1$'s redundant units in algorithm E. $u_2$ is the other one unit in the same 2-unit layer with $u_1$. When we get the current goal generation of a last chosen unit in algorithm E1, $u_2$'s redundant units have been deleted. The new 2-unit layers whose third deleted units are $u_2$'s redundant units are $u_2$'s son 2-unit layers. When the goal generation is a final leaf's goal generation, we call $\boldsymbol{u_1}$ **the final leaf's goal generation's father** and call $\boldsymbol{u_2}$ the final leaf's goal generation's **father in law**.

Now this is the key: in each generation, there is at least one layer whose third unit is a correct unit and whose first two units are wrong units. We cannot directly get such layers, but we can determine that at least one last chosen unit and its ancestor chosen units are correct units. Consider this case. We calculate a unit $u$ in algorithm E and delete its redundant units. $v_1$ and $v_2$ are two wrong units in one of $u$'s son 2-unit layers ($v_1$ is the second one). In these son 2-unit layers, $p_1$ and $p_2$ are two part paths which only contain correct units. $p_1$ destroys $p_2$ but before $u$'s redundant units were deleted, $p_1$ did not destroy $p_2$, that is, the contradiction between $p_1$ and $p_2$ is caused by $v_1$ and $v_2$ ($v_1$ destroys $p_2$ and $v_2$ destroys $p_1$). After $u$ was calculated, we calculate $u$'s one son unit in $p_1$. All these we know though we do not know which units are correct units and which are wrong units. Later when we calculate chosen units in algorithm E1, if a chosen unit $c$ in a lower generation is exactly destroyed by $v_1$ and also exactly destroyed by the units in the same 2-unit layers with $p_2$ (units in $p_2$ are the first units), or is not exactly destroyed by $v_1$ but only exactly destroyed by the units in the same 2-unit layers with $p_2$, then $c$ has two possible father chosen units: one is the third unit in $v_1$'s layer, the others are the third units in $p_2$'s layers. We only choose one of them, that is, discard the other. How to choose? Firstly keep the third unit in $v_1$'s layer and all the descendant chosen units (only keep these chosen units in their layers, delete the first two units), in $p_1$'s and $p_2$'s layers, only keep $p_1$ and $p_2$, recover all other layers' three units in the current goal generation and the layers under it, and also for the calculated unit whose son 2-unit layers are the supposed current goal generation, delete this unit. If the standard checking tree property holds (do the additional job for checking), we choose this third unit as $c$'s father chosen unit (do not mind whether they are correct units or wrong units, note that correct units only exactly destroy wrong units). We call this **the changing father chosen unit rule**. If the property does not hold, we let the third units in $p_2$'s layers as $c$'s father chosen units. This is because if all the chosen units are correct units and the goal generation contains the first occurred wrong 2-unit layer, the standard checking tree property must hold and if the chosen units contain wrong units they must destroy some correct units so that the property does not hold (see the method to get a key part path in the following). This rule can also guarantee that for each successful calling of algorithm E1, we can get that at least one last chosen unit with its ancestor chosen units are correct units (when the current goal generation contains the first occurred wrong 2-unit layer). We call this **the one correct last chosen unit property**. Note: 1) $v_1$ destroys $p_2$ and $v_2$ destroys $p_1$. Such destroying is not caused by other destroying parallel pairs in 2-unit layers. 2) $v_1$ and $v_2$ may not be in the same layer. We consider the third unit in each layer as possible father chosen unit respectively. 3) the first two units in $p_1$'s or in $p_2$'s layers may also be as another $v_1$ and $v_2$. Also consider the non-friend last chosen unit rule (see next).

After we take a unit as the last chosen unit, suppose in each chosen unit $x$'s father chosen units' layers, the first units (each layer's first unit) destroy $x$. So we have to consider the second units in chosen units' layers. For this we also may have to change a chosen unit's father chosen units. Consider this case: let $c_1$ be a chosen unit in a generation (this is the lower generation) and suppose it is a correct unit. In the generation just above this generation, there are two parallel part paths in the first two units. We call this generation the upper generation. If the last real long path contains $c_1$ and a part of one of the two parallel part paths, $c_1$'s father chosen units are the third units in whose layers some units exactly destroy $c_1$. If the last real long path contains $c_1$ and a part of the other one part path, $c_1$'s father chosen units would be other third units. So we have to check each third unit in the upper generation. We firstly try the third units in whose layers some units exactly destroy $c_1$. We call the above case and this case both **the changing father chosen units case**.

Consider a last chosen unit $ld$ with its ancestor chosen units. Suppose that in algorithm E if we always choose correct units to calculate, we get a generation in which $w_1$ and $w_2$ are two parallel part paths (the first two units) and $w_3$ is the third unit set. $dv_1$, $dv_2$, $dv_3$ are three units in one layer in this generation. The third units in this generation have been deleted and the standard checking tree property holds. Then if we only delete $dv_1$'s redundant units (third redundant units), we get $dv_1$'s son 2-unit layers which are a second kind last generation. But if we delete $dv_2$'s redundant units the son layers are not a last generation. Suppose that in the 3-unit layers under the second kind last generation some units are destroyed by $ld$ and its ancestor chosen units. If the destroyed units are correct units and the chosen units are wrong units, because $dv_2$'s son 2-unit layers are not a last generation, the standard checking tree property still holds even do the additional job for checking. We can solve this problem in this way: only consider two units in each layer, that is, do not consider the units in this part path which contains $dv_2$. Then in each other generation, we also only consider two units in each layer in the same way.

In algorithm E, suppose when the standard checking tree property holds, there are two parallel 2SAT paths over the 2-unit layers (or two parallel part paths over some 2-unit layers). Then we choose a unit to calculate and get its son generation. Then calculate a son unit, and then son's son, until get a last generation. We call all these layers **a two parallel 2SAT paths structure**.

Note **this structure's four properties**: 1) two or more structures may overlap, but each structure is independent, that is, we can independently get each structure; 2) all structures share the same long paths; 3) in a structure, each generation is independent, that is, we can separate each generation from other generations; 4) deleting trouble units and then letting the standard checking tree property hold do not affect the above three properties and the three or more long paths property.

Consider the two two parallel 2SAT paths structures. One contains the chosen units (structure 1) and the other one contains a second kind last generation (structure 2). These two structures may overlap and the second kind last generation may disappear. This does not affect our method. No matter the two structures overlap or do not overlap, wrong chosen units in structure 1 (not in structure 2) definitely destroy some correct units in the last 3-unit layers under a structure 2. Note that such second kind last generations may be more than one. This does not affect our method.

**The key is to find** the last 3-unit layers and the second kind last generation. Based on the key part path property, the standard checking tree property and the three or more long paths property, when the current goal generation contains the first occurred wrong 2-unit layer (delete the current goal generation's father unit), a wrong chosen unit definitely destroys some correct units in some last 3-unit layers unless it is a third redundant unit in structure 2. If it is a third redundant unit in structure 2, it must destroy some correct units in structure 2. We always have some chosen units which are not in structure 2 (if structure 1 and structure 2 utterly overlap, we can easily find the key part path and the chosen units which are in the same one long path with the key part path). Two last chosen units may be in different real long paths. So if we delete two or more last chosen units with their ancestor chosen units' redundant units, this may make that a part path over a second kind last generation does not have the same one useful 2SAT path. Thus by the additional job for checking, we can determine a second kind last generation and the last 3-unit layers. Check that the three or more long paths property does not fit some part paths over the last 3-unit layers. Then we get a part path over the second kind last generation and the last 3-unit layers and this part path does not contain the third redundant units in the second kind last generation. The three or more long paths property does not fit the part of this part path over the last 3-unit layers (or this part is surely in a real long path). If the current goal generation contains the first occurred wrong 2-unit layer and its father unit was deleted, we always have such a key part path. Units in this part path are also as the final leaf's good units when this job finishes. If a chosen unit is a third redundant unit in structure 2, based on the three or more long paths property and the part path we got over the second kind last generation and the last 3-unit layers, we can separate the correct unit from other two units in this unit's layer.

Based on the above we have **the method to get a key part path**:

For a last chosen unit and its ancestor chosen units, if after we delete their redundant units and recover some units in other layers, we firstly try to get a key part path if we have not got. When we calculate chosen units and delete their redundant units, and just when we begin to check the standard checking tree property, if we can find such a 3-unit layer $t_1$, $t_2$, $t_3$ and such a part path *pp*: $t_3$ is destroyed by the chosen units (if both $t_2$ and $t_3$ are destroyed by the chosen units, this does not affect our method), units in *pp* do not have the same one useful 2SAT path (that is, some units in it destroy $t_1$ and some other units in it destroy $t_2$), there is another part path *ap* which is over such 3-unit layers that contain units destroyed by *pp*, *ap* is surely in one real long path, *ap* contains the unit $t_3$, and *ap* does not destroy *pp*, then *pp* together with *ap* is **a key part path**. **Note that** *ap* may be in structure 1, but this does not affect our method. If we cannot find the key part path, that is, the last chosen unit and its ancestor chosen units do not destroy enough correct units, then the last chosen unit and its ancestor chosen units are correct units and we still can continue to calculate. If for more than one such case we can get more than one key part path, we only choose one case and set units in this case's key part path as correct units. So we need to do a job to check whether the key part path belongs to the current goal generation's descendants. We only need such a key part path and neglect other possible key part paths.

In order to get the key part path, for the last generation's two parallel part paths over the first two units in structure 1, each time we only consider one (that is, temporarily delete the other one).

The **job to check** whether the key part path belongs to the current goal generation's descendants. After we got a key part path, we temporarily delete some units in the last 3-unit layers so that when the standard checking tree property holds, the key part path (the part over the second kind last generation) does not have the same one useful 2-SAT path. Then by the additional job for checking, we get a part path over the father generation which also does not have the same one useful 2SAT path. Then in father's father generation, get such a part path. And so on. This part path over the father generation destroys the third redundant units in the son generation. When we do the additional job for checking, we only consider two units in each layer, that is, only consider the part path and the third redundant units so that in the father generation we can get such a part path which also does not have the same one useful 2SAT path (if consider all three units, this part path may have the same one useful 2SAT path). At last, if we get a father generation which is the current goal generation, we say that **the key part path belongs to the current goal generation's descendants**. This **job is to check** whether the key part path belongs to the current goal generation's descendants. Note that in each generation, based on the three or more long paths property, the third redundant units are separated into three or more parts and each part is in a different long path. Based on this we can get the third units. If we cannot separate the third units, that is, other units also have the third units' property, this means that each of some part paths over the first two units is in a real long path and thus the chosen units are always correct units (if not so, based on the additional job for checking (consider all three units), there is no part path over the first two units which has the same one useful 2SAT path). So we do not need to mind this case.

In this job, we can get **an extended key part path**. It is the key part path together with every part path over each generation in this job which are in the same one long path. Later we calculate chosen units based on the extended key part path.

Note that the basic logic for the job to check whether the key part path belongs to the current goal generation's descendants is that in structure 2 there are two parallel part paths in the first two units in each generation (these two parallel part paths over the first two units or part of the first two units in each generation), and if there are some intersection part paths between these two parallel part paths (that is, such a part path consists of two parts, one part is in the first parallel part path and the other one part is in the second parallel part path), then whether such an intersection part path is in a real long path, or, when we do the additional job for checking as calculating a last chosen unit's goal generation while this last chosen unit and its ancestor chosen units contain wrong units, the standard checking tree property does not hold (when the current goal generation contains the first occurred wrong 2-unit layer and the key part path belongs to the current goal generation's descendants and also the wrong chosen units (together with one of the parallel part paths in the last generation in structure 1) make part of the key part path not have the same one useful 2SAT path). We call this **the parallel part paths property 1**. This property is the key logic for the whole algorithm.

If the current goal generation contains the first occurred wrong 2-unit layer and the key part path belongs to the current goal generation's descendants, we call this key part path **a good key part path**. A good key part path is in a last real long path. If a key part path is not in a last real long path, it is **a bad key part path**. A bad key part path certainly destroys a good key part path. When the standard checking tree property holds, if the key part path is a good key part path, then all the chosen units are in the same one real long path, that is, this case cannot happen: the chosen units are not in the same one real long path and the three or more long paths property fits them. In this way, if a third chosen unit is in structure 2 and is not a correct unit, it may not destroy the key part path and may also make the standard checking tree property hold, but in each generation we always can get such third chosen units which are correct units.

In algorithm E1, when we delete the redundant units of a last chosen unit and its ancestor chosen units and check the standard checking tree property we firstly have to try to get a key part path unless we have got this key part path. For each current goal generation, we need to get one key part path. The key part path has to belong to the current goal generation's descendants. We calculate the key part path by the additional job for checking and by the method as stated above. If we cannot get a key part path, still calculate as usual. If we have got a key part path, only keep this key part path's units in its layers and then calculate from the start in the same way.

In the mean time, delete the unit $u_1$ ($u_1$ is the current goal generation's father), only keep the last chosen unit with its ancestor chosen units in their layers, and also recover other layers' three units in the generations not higher than the current goal generation.

**So in the whole algorithm E1, when we check the standard checking tree property, together with this job we also have to consider the (extended) key part path**.

A summary of the algorithm E1:

1) Each time we suppose a goal generation and we call it the supposed current goal generation, or the current goal generation. For each last chosen unit, we only calculate whether it has such goal generation. We firstly suppose the generation just above the last generation as the current goal generation. Then if it failed, we suppose the generation just above it as the current goal generation, and so on. If the current goal generation changed, recalculate each last chosen unit.

2) In a generation which is not higher than the current goal generation and in which each last chosen unit's ancestor chosen units have been got, for any two last chosen units, when we only keep the two last chosen units with their ancestor chosen units in their layers and keep other layers' three units from the last generation to this generation, from this generation to the current goal generation keep all layers' three units, and delete the unit $u_1$ ($u_1$ is the current goal generation's father), then if the standard checking tree property holds, we call the two last chosen units **friend last chosen units**. If the property does not hold, we call them **non-friend last chosen units**. At first, just in the last generation, we calculate each last chosen unit's friend last chosen units in this way. Then, in each higher generation, we update each last chosen unit's friend last chosen units.

3) At first, for each last chosen unit, if we delete the first two units in its layer, delete the unit $u_1$ and recover all other layers' three units in the generations under $u_1$, then if the standard checking tree property does not hold, we do not calculate such last chosen units (temporarily delete them when calculate other last chosen units).

4) When we calculate a last chosen unit's ancestor chosen units in a generation and check the standard checking tree property, we have to temporarily delete this last chosen unit's non-friend last chosen units with their ancestor chosen units up to the generation just under this generation and then check the property. Also delete the third units which are not any last chosen unit's ancestors in the generations under this generation. We call this **the non-friend last chosen unit rule**. Note that some units which are not correct units may not destroy any correct units under its layer so that we may take them as correct units. The non-friend last chosen unit rule can solve this problem. Also note that when recovering the third unit, the additional contradiction pairs caused by the first two units when the third unit was deleted have to be recovered to no contradiction.

5) How to change a chosen unit's father chosen units? Consider the changing father chosen units case (see above). Consider the chosen unit $c_1$ in the lower generation. From the lower generation to the last generation, calculate each last chosen unit's friend last chosen units. In these generations, delete all last chosen units with their ancestor chosen units which are non-friend last chosen units of the last chosen unit which is $c_1$'s descendant unit and also delete the third units which are not any last chosen unit's ancestors (keep the first two units in these layers). Then take each third unit in the upper generation as $c_1$'s father chosen unit respectively. Only keep this last chosen unit with its ancestor chosen units up to $c_1$'s such father chosen unit in these units' layers, delete the unit $u_1$ ($u_1$ is the current goal generation's father), and recover all other layers' three units from the current goal generation to the last generation, then if the standard checking tree property holds, this father chosen unit is really $c_1$'s father chosen unit. Otherwise, it is not. If such father chosen units are more than one (only keep them in their layers the standard checking tree property holds), they are all $c_1$'s father chosen units. We firstly try to take the third units in whose layers some units exactly destroy $c_1$ as $c_1$'s father chosen units. If succeed, do not calculate other possible father chosen units.

6) When we calculate units in structure 1 (stated above) in algorithm E, for a layer in which the third redundant unit is a correct unit, if this layer is not in structure 2 as well as the last 3-unit layers, then it must have a son layer which is not in structure 2 and the third unit in this son layer is a correct unit. Then in each later generation, there is at least one layer whose third deleted unit is a correct unit and this one unit in this layer can make the standard checking tree property hold. This is because a wrong unit destroys a correct unit and this correct unit cannot be in structure 2. We call this **the son layer property**. If all third deleted correct units are in structure 2 and then the standard checking tree property does not hold, we can get the (extended) key part path by the above method and then we can get some chosen units which are correct units. If some deleted correct units are not in structure 2, based on the son layer property we can get a last chosen unit with its some ancestor chosen units which are correct units.

7) In algorithm E1, when we delete the redundant units of a last chosen unit and its ancestor chosen units and check the standard checking tree property, we firstly always have to try to get a (extended) key part path unless we have got this key part path. For each current goal generation, we need to get one key part path. We calculate the key part path by the method to get a key part path. If we have not got a key part path, only keep the last chosen unit and its ancestor chosen units in their layers, delete the non-friend chosen units of this last chosen unit, and for other layers not higher than the current goal generation recover all 3 units in each layer. Then if the standard checking tree property holds and if the current goal generation contains the first occurred wrong 2-unit layer, the last chosen unit and its ancestor chosen units are correct units. If the property does not hold, the chosen units contain wrong units. If we have got a key part path, only keep the (extended) key part path's units in its layers and then calculate as above. Units in the extended key part path are also as the final leaf's good units. Note that if there are no key part paths, or there are some key part paths but no key part paths belong to the current goal generation's descendants, then we treat this case as no key part paths and thus directly calculate the chosen units.

8) In the last generation of structure 1, in the first two units, there are two parallel part paths $tp_{11}$ and $tp_{12}$ and there are no intersection part paths between these two parallel part paths. Also in the second kind last generation of structure 2, in the first two units, there are two such parallel part paths (if not, see the parallel part paths property 1) $tp_{21}$ and $tp_{22}$. When we calculate the key part path, we delete one of the $tp_{11}$ and $tp_{12}$. Then after we got the key part path, we only keep one of $tp_{11}$ and $tp_{12}$ for later calculating chosen units. The key part path contains one of $tp_{21}$ and $tp_{22}$. These two parallel part paths let us separate the third units and the first two units. Note that structure 1 and structure 2 may share some layers, but this does not affect our method. We can easily separate these layers from other layers. Based on the parallel part paths property 1, in each generation we can separate the third units from this generation. For special case, the first two units do not contain parallel part paths, we still can separate the third units and the first two units or we can check the standard checking tree property does not hold. Based on this, after we got the key part path, we can get a part path over the second kind last generation's father generation which is in the same one long path with the key part path. Then get such a part path over the father's father generation. And so on. Though there may be more than one part path over different generations which destroy the third units in one generation, but this does not affect that we get the part path over the father generation. Also, if the structure 1 and the structure 2 overlap, we can directly get a key part path (contain some third units) in the structure 1 and some layers under the last generation. Then check whether the key part path belongs to the current goal generation's descendants. **Note tha**t when we calculate a last chosen unit and its ancestor chosen units, especially when we calculate the key part path, each time we only keep one of the two parallel part paths over each generation in structure 1.

9) Note that in algorithm E, after the generation which contains the first occurred wrong 2-unit layer, considering the structure overlapping, we may still calculate a correct unit and this correct unit's redundant units are all wrong units. If then we always do so until get the last generation, we can directly get the key part path over the last generation and some layers under the last generation. Then in algorithm E1 we check whether the key part path belongs to the current goal generation. If it belongs, we get the extended key part path. Also, when we calculate chosen units in algorithm E1, if in a generation which contains a part path contained in the extended key part path, do not calculate this generation's chosen units. In some layers under the last generation in algorithm E, we can get some part paths each of which is surely in a long path, but such layers may be in the upper 2-unit layers (the third units were deleted). Also under the last generation, there may be another generation whose third units are destroyed by some units in the last generation. These do not affect our method.

**Note that the parallel part paths** over the first two units in each generation (the generations not higher than the first occurred wrong 2-unit layer) without intersection part paths between them is the key for the 2SAT logic. **Based on this**, we can separate the third units from the first two units in structure 2, and also we can separate different generations. In structure 2, if there are some intersection part paths between these two parallel part paths and this does not affect that we separate the third units from the first two units, this does not matter, and if this affects that we separate the third units, see the parallel part paths property 1.

Usually in each generation in structure 1 (from the generation which contains the first occurred wrong 2-unit layer to the last generation), there are two parallel part paths over the first two units without intersection part paths between them. The three or more long paths property fits each of the two parallel part paths. If not so for such a generation, that is, there is at least one intersection path between the two parallel part paths, then such a intersection part path must be in a long path. This is a contradiction. Note that there is a special case for some such generation that there are intersection part paths between these two parallel part paths but each intersection part path is not in a long path. This special case has this property: each parallel part path and each intersection part path have the same three or more long paths character (that is, each such part path can be separated into three or more shorter part paths in the same way by layers). We call this case **the special parallel part paths case** and call this property **the special parallel part paths property**. Due to this property, **this special case does not affect our method**. So if in structure 2 there is such a special case, then when we get the last generation in algorithm E and do the additional job for checking, the standard checking tree property must not hold (as stated formerly). Also for this special case, calculating correct chosen units in algorithm E1 is easy (does not need to consider the changing father chosen units case). For this special case, we need to consider the hidden destroying parallel pairs when calculating chosen units. Even for the special case, there are still parallel part paths as stated above in the last generation (when consider the layers in which the third units are destroyed by hidden destroying parallel pairs). We call all these **the parallel part paths property 2**.

Based on the changing father chosen units case and the key part path property, a correct last chosen unit may not be in the last generation, but in a higher generation (or, as stated above, in a hidden destroying parallel pair's layer). This does not affect our method. We calculate such last chosen units in the same way.

Also, in algorithm E, when we delete a unit *ul*'s redundant units, get the last generation and there are no current real long paths, the standard checking tree property may still hold due to hidden destroying parallel pairs. Then we can find a unit *u* in the last generation so that if we delete *u*'s redundant units in the rest 3-unit layers, there is a part path over the new 2-unit layers (*u*'s son 2-unit layers) which is surely in a real long path. Then we do the algorithm E1 to calculate the final leaf. We call this last generation **a special last generation**. Note that the last generation only contains *ul*'s redundant units' layers, these redundant units are destroyed by *ul* before they are deleted.

In algorithm E, from the generation which contains the first occurred wrong 2-unit layer, each generation contains such parallel part paths $g_1$ and $g_2$ with utterly different units, there are no intersection part paths between these parallel part paths and between $g_1$ (or $g_2$) and the third deleted units there are three or more intersection part paths. At least one of these intersection part paths is in the last real long path (if it exists). In each generation, if we choose one unit in $g_1$ or $g_2$ to calculate, when we get to the last generation and the third units in the last generation are deleted, the standard checking tree property does not hold. And then we can calculate the key part path and the chosen units in algorithm E1. But we may calculate a unit which is not in $g_1$ and $g_2$ but in another parallel part paths. How to solve this problem? In the last generation (we do not know the last generation, but we can try each generation; this time the last generation may not mean that the standard checking tree property does not hold, but means that we can get the next generation over which there is a part path that is surely in a long path and based on this we can calculate the key part path or correct units), we firstly check whether one of the parallel part paths or whether all the third deleted units in this generation are in a last real long path by trying to find the key part path (in the same logic as above). If so, we can get correct units when the goal generation contains the first occurred wrong 2-unit layer. If not, we calculate the key part path as usual (that is, as stated formerly). We call this **the hidden destroying parallel pair rule**. Also note the special parallel part paths case as stated above. In this special case, there are some intersection part paths between $g_1$ and $g_2$. This does not affect our method. Also note the shorter single part path rule in algorithm E.

At last, after all layers in the last generation have been calculated, maybe more than one last chosen layer has the same goal generation or maybe some chosen units do not have goal generations (a goal generation should be lower than the first generation). Now we have one or more last chosen units which have the current goal generations (if not, we suppose the generation just above the current goal generation as the new current goal generation and do again).

We call the goal generation which contains the first occurred wrong 2-unit layer **the correct goal generation**. If a last chosen unit's goal generation is the correct goal generation, then it and its ancestor chosen units are correct units and also units in the key part path which belongs to the current goal generation's descendants are correct units. We call this **the correct chosen units property**. For this case, if the chosen units are wrong units, they must destroy some correct units and let some 2-unit layers only contain wrong units. By the three or more long paths property, especially when we only keep the key part path's units in their layers, the standard checking tree property must not hold. For the correct chosen units property, also consider the one correct last chosen unit property. The correct chosen units property is the key for this algorithm. It is hard to understand.

Note: if a last chosen unit's goal generation is the correct goal generation, its ancestor chosen units must be correct units. But two such last chosen units with their ancestor chosen units may not be in the same one real long path. So each time we only choose one last chosen unit as the final leaf. Then other last chosen units or some of them which have the same one goal generation with their ancestor chosen units may not be correct units (we only concern one real long path).

If in one generation we can get a part path over the whole 2-unit layers which only contains finished units, this means there is a current real long path and we have got the goal generation. We call it **the second kind goal generation**. Let *f* be the final leaf we have got in algorithm E1 and *f*'s goal generation is $u_1$'s son 2-unit layers. $u_2$ is the unit in the same 2-unit layer with $u_1$. Note that if $u_1$'s son 2-unit layers contain the first occurred wrong 2-unit layer, for $u_2$'s any one descendant correct unit, if we calculate it as a last chosen unit in a new calling of algorithm E1, its goal generation is lower than $u_2$'s son 2-unit layers, or we get the second kind goal generation. But if some correct units in $u_2$'s son 2-unit layers were deleted, we would not get the second kind goal generation. If $u_1$'s son 2-unit layers do not contain the first occurred wrong 2-unit layer, for $u_2$'s any one descendant correct unit, if we calculate it as a last chosen unit in another calling algorithm E1, whether its goal generation is lower than $u_2$'s son 2-unit layers, or is higher than $u_2$'s son 2-unit layers. This time we cannot get the second kind goal generation. If its goal generation is higher than $u_2$'s son 2-unit layers, we have to combine this new final leaf and $f$ into one final leaf. These two final leaves **share ancestor chosen units in upper generations which are higher than $u_1$'s son 2-unit layers**, so this combination is one to one. Under $f$, we have to try to calculate all possible units in algorithm E to let the new final leaf's goal generation be lower than $f$ or to get the second kind goal generation. If we cannot get so, then we combine two final leaves into one final leaf whose goal generation is higher. If we combine two final leaves, $f$'s goal generation has to be changed to higher. Because this combination is one to one, a last chosen unit in the new algorithm E1 may not be combined with $f$, but with another last chosen unit which was calculated at the same calling of algorithm E1 with $f$. If it is not combined with $f$, the first two units in $f$'s good units' layers have to be recovered. Then every two combined last chosen units have a new goal generation. We choose such combined two last chosen unis whose new goal generation is the lowest one. We will discuss this combination next. We call this **the one to one final leaf combination rule**. In this rule, note the changing father chosen unit rule, the one correct last chosen unit property and the correct chosen units property.

Note that for the one to one final leaf combination to combine two last chosen units, the two last chosen units share some ancestor chosen units and also share the same one key part path. If the key part path changes, each last chosen unit's ancestor chosen units may also change.

Suppose the current goal generation contains the first occurred wrong 2-unit layer, that is, $u_1$ is the first calculated wrong unit. Suppose in algorithm E, when we calculate $u_1$, we do not calculate $u_1$ but calculate the correct unit $u_2$ in the same layer. Consider $u_2$'s son 2-unit layers. Suppose each time in each generation we always take a correct unit to calculate. Delete its redundant units and let the standard checking tree property hold. Then calculate the next generation, then next again. Then at last we surely can get a second kind goal generation.

If the current goal generation is not lower than a former final leaf's goal generation, we call the former final leaf **a failed final leaf**.

For a failed final leaf, its good units cannot be as a new final leaf's good units unless the new final leaf has more good units and its goal generation as well as the generations under this goal generation cannot be as a goal generation later. We call this **the goal generation discarding rule**. This rule can guarantee that we do not repeat to calculate wrong units. Each time when we get a final leaf, we put the final leaf with its good units under $u_2$ and over $u_2$'s son 2-unit layers.

Note that at any time, the correct goal generation (the generation which contains the first occurred wrong 2-unit layer) is only one. Suppose that we firstly get a final leaf $f_1$ and then under $f_1$ we get a new final leaf $f_2$ (in another time to do algorithm E1). Suppose both are correct leaves. If $f_1$'s goal generation is the correct one which contains the first occurred wrong 2-unit layer, then for $f_2$, its correct goal generation is under $f_1$ which contains the new first occurred wrong 2-unit layer. If $f_1$'s goal generation is wrong, that is, the correct goal generation is a higher generation, then for both $f_1$ and $f_2$, the correct goal generation is this higher generation (that is, the same one). We call these **the one correct goal generation property**. Note that if the goal generation is a higher generation, we can combine $f_1$ and $f_2$ into one final leaf, that is, only let $f_1$ as the final leaf, the good units of $f_1$ and $f_2$, including the increased good units in higher generations, are $f_1$'s good units. We also combine more two final leaves, one was got at the same time with $f_1$ and the other one was got with $f_2$, into one final leaf in this way, one to one (two final leaves share ancestor chosen units in upper generations). In this way, a correct leaf is certainly combined to another correct leaf. As stated above, we call this **the one to one final leaf combination rule**. This rule, the one correct goal generation property, let us not do exponential combinations. Now suppose $f_1$'s goal generation is not the correct one. In a new calling algorithm E1, we calculate the goal generations of $f_2$ and other last chosen units which are calculated together with $f_2$. We get the final leaf $f_2$. Then we may still get some final leaves under $f_2$. But at last we cannot get more final leaves under $f_2$ and also cannot get the second kind goal generation. Then we have to combine two final leaves into one final leaf by the one to one final leaf combination rule. We firstly try to keep $f_1$ and its goal generation unchanged. But we cannot get a goal generation under this condition. Then we do not need to try another one last chosen unit which had the same goal generation with $f_1$ when $f_1$ was calculated (we remember all such last chosen units) but directly change to a higher goal generation, because of the correct chosen units property. We set a higher generation as $f_1$'s current goal generation and recalculate the new goal generations of $f_1$ and other last chosen units which were calculated at the same time with $f_1$. This time, $f_1$ and $f_2$ have the same correct goal generation. So we suppose a new higher goal generation as their current goal generation. Combine them together to calculate the new goal generation (by getting ancestor chosen units and checking the standard checking tree property). We also combine more two final leaves, one was got at the same time with $f_1$ and the other one was got with $f_2$, to calculate their goal generation (final leaves under $f_2$ have been combined by the one to one rule). Note that as stated above, the combination is **one to one**. After the goal generation changed, we recalculate each last chosen unit.

A final leaf's each good unit can also be as a final leaf and these good units are its good units. A final leaf's good units will increase when the supposed current goal generation becomes higher. After we got a final leaf with its good units, later we may change this final leaf to another final leaf (see the above). We also may keep this final leaf but change the goal generation to a higher generation (redo the algorithm E1 under the same condition as the time when we got the final leaf) and thus the good units increase. We do not let a final leaf disappear but without a new final leaf replacing it. We call this **the no decreasing rule for final leaves**. This rule and the one to one final leaf combination rule let us avoid repeating to calculate final leaves.

A failed final leaf with its good units cannot be as a final leaf and good units later unless the new goal generation is higher and with more good units. We call this **the failed final leaf rule**.

If we can get a final leaf, we say that the algorithm E1 **succeeds**. If there is no final leaf for all possible generations, the algorithm E1 **fails** and there is no long path this time.

In the above way, if the current goal generation contains the first occurred wrong 2-unit layer, we always can get a last chosen unit and its ancestor chosen units which are correct units and then these correct units are always kept and do not change. If we get a last chosen unit which is a wrong unit, we always have the opportunity to get a correct one with a higher goal generation and in the mean time we do not need to do exponential combinations.

Now for the final leaf, we remember its good units. We treat it as a leaf of kind 1 (now do not mind the leaf $l$ in algorithm E). The first two units in the same layer with the leaf are also this leaf's destroying parallel pair and **the leaf is also a good unit of itself**. If the last real long paths exist, at last we can get one last chosen unit and its ancestor chosen units which are correct units. Also units in the key part path are as the final leaf's good units.

If the current goal generation contains the first occurred wrong 2-unit layers, we can get a final leaf with its good units which are correct units. But the current goal generation may be lower than the generation which contains the first occurred wrong 2-unit layers. If so, the final leaf and its good units are not certainly correct units. But we can update it later and get a higher goal generation. Note that in different times to do algorithm E1, some calculated chosen units may be the same. But this does not affect our method.

**The key problem is,** for different final leaves with their good units, how to get a combination of only correct final leaves whose good units are correct units in polynomial time?

Note that in one calling of the algorithm E1, may be more than one last chosen unit and its ancestor chosen units are correct units. If we keep two correct last chosen units and their ancestor chosen units in their layers and delete other two units in these layers (when the goal generation contains the first occurred wrong unit and no correct units were deleted), the standard checking tree property may hold (the property may not hold, because these two correct last chosen units may not be in the same one real long path. If so and if we only keep one last chosen unit with its good units in their layers, the other one last chosen unit with its good units would become wrong units). Because we cannot do exponential combinations, this is a problem. But this problem can be easily solved. For each calling of algorithm E1, we only care the first correct final leaf. If in each calling of algorithm E1, the first final leaf is a correct leaf, other correct leaves in the same one calling of algorithm E1 can be got automatically (because in these correct units' layers, if all three units are kept (not deleted), this does not matter and if some units are destroyed and deleted by the first correct leaf, this still does not matter due to that the destroyed units are wrong units). We call this **the first correct leaf property**.

Based on the above properties and rules, at last we can get all the first correct leaves in each calling of algorithm E1.

Consider the one to one final leaf combination rule. This rule means that for *k* (or more) last chosen units in one calling of algorithm E1, and *k* (or more) last chosen units in another one calling of algorithm E1, there are *k* (only *k*) pairs last chosen units. Each pair share ancestor chosen units in some upper generations. Note that for the *k* pairs of last chosen units, each pair contains two sides. Each side contains one or more last chosen units. The two sides share ancestor chosen units in upper generations not lower than the former (failed) goal generation. When we do the algorithm E1 and try to calculate a last chosen unit *c*, for an old final leaf *f* and its good units which were got in a former calling of algorithm E1, if *c'*s goal generation is lower than *f* and thus is not affected by *f*, we can continue; if *c'*s goal generation is higher than *f*, we handle this by one to one final leaf combination rule; if *c* is a correct unit, *f* is a wrong unit, *c*'s correct goal generation is affected by *f* and thus we cannot get *c*'s correct goal generation, then by the one to one final leaf combination rule, we can change *f* to a correct leaf which was calculated in the same calling of algorithm E1 with *f* and thus *c* can get its correct goal generation (or get the second kind goal generation).

After we got a final leaf *f*, under this final leaf we calculate units according to the calculation order. If we cannot get a finished unit, we try to get a new final leaf whose goal generation is lower than *f*. If after we calculated all possible units (only need to calculate according to the calculation order) and cannot get such a leaf and also cannot get a second kind goal generation under *f*, then we can get a new final leaf with the goal generation not lower than *f*. Then we say *f* **a failed final leaf**. A failed final leaf (include any one of its good units) cannot be as a final leaf again unless with more good units. This is the key for polynomial. But how get a new final leaf with the goal generation not lower than *f*? After we got the final leaf *f*, when we firstly do the algorithm E1, we can get this new final leaf. In the mean time, note the one to one final leaf combination rule. We choose the combined two last chosen units whose goal generation is the lowest as the new final leaf (one of the two as new leaf). Note that after we combined two final leaves, then under another former final leaf, we still keep the rule: each unit is calculated one time for this final leaf (in algorithm E); if a unit was calculated for this final leaf, it cannot be calculated again for this leaf. And this former final leaf also may be combined in the same way and same logic.

Note that in the algorithm E, each time we choose one layer in the 2-unit layers and choose one unit in this layer to calculate. Delete this unit's redundant units and let the standard checking tree property hold. Then in the new 2-unit layers, we also choose one layer and one unit to calculate in this way. If after the unit's redundant units were deleted, the standard checking tree property does not hold, we recover the redundant units just deleted and then calculate the other one unit in the same one layer, that is, in such a 2-unit layer, we always choose such a unit to calculate when whose redundant units were deleted the standard checking tree property holds. We call such a unit **a success unit**. Until, we get such a new 2-unit layer, when the redundant units of any one unit in this layer were deleted, the standard checking tree property does not hold. We choose each one of these two units to calculate and to get the last generation. Then we calculate each last chosen unit by the algorithm E1. We call this **the success unit rule**.

When we finished the algorithm E1 and got the final leaf with its good units, some layers changed and we return to the algorithm E. Note that at this time, the exact contradiction pairs are the contradiction pairs just before the third unit in the final leaf's goal generation's father's layer was deleted.

In algorithm E1, we get the final leaf with its good units. Each time we only get one final leaf with its good units. Note that at this time we only see this final leaf's good units as **the units we have calculated**. Based on this, **for each final leaf with the same good units we calculate a unit at most one time (in algorithm E)**.

Finally, though each time we only need to get a correct last chosen unit and its correct good units, but we can see: in this algorithm E1, each time we have a supposed current goal generation. If this supposed goal generation contains the first occurred wrong 2-unit layer, then all the chosen units whose goal generations are this generation are correct units, and each correct last chosen unit must have this goal generation. But note that two correct last chosen units may not be in the same one real long path and if so when we choose one last chosen unit as the final leaf, the other one last chosen unit with its good units would become wrong units. If this goal generation does not contain the first occurred wrong 2-unit layer, we may get wrong result. But this does not matter and we only need that when it does contain, we get the correct result.

PROOF.

Based on the standard checking tree property and the three or more long paths property, we can get the algorithm E1.

**The key for the polynomial time is**: each time we suppose a current goal generation. For each last chosen unit, we calculate it independently whether it has this goal generation and do not need to do a lot of combinations of different last chosen units with their ancestor chosen units. If this current goal generation fails, it will not be as a current goal generation again and the new current goal generation is higher (in one algorithm E and for all algorithm E1 in this algorithm E).

There is at least one layer in the current first generation of wrong 2-unit layers which contains only wrong units. Then in each descendant generation, this layer's descendant layers must contain some layers which only contain wrong units (the first two units) when hidden destroying parallel pairs are considered (if possible, we always let lower layers contain wrong units and higher layers contain correct units, see the lower wrong 2-unit layer rule).

So in the current last generation of wrong 2-unit layers (including hidden destroying parallel pairs' descendants), we always can get such a layer (it exists) in which the first two units are wrong units and the first two units in its all ancestor layers up to the current first generation of wrong 2-unit layers are wrong units, and thus when we recover these layers' third units and delete the first two units and other layers' three units in these generations are kept, the standard checking tree property must hold.

When the standard checking tree property holds, there are two possible cases. Case 1: there is one or more real long paths over the current tree and these real long paths make the property hold. Case 2: there are no real long paths now, but based on the three or more long paths property, the standard checking tree property holds.

**Note that** in the current first generation of wrong 2-unit layers, there is at least one layer whose first two units are wrong units and in each descendant generation, this layer's descendant layers contain at least one layer whose first two units are wrong units (considering hidden destroying parallel pairs). Also such a descendant layer's father layers only contain wrong units (the first two units), because if not so, the current first generation must not be the first generation of wrong 2-unit layers, that is, each layer's first two units in this generation must contain at least one correct unit. This is the key for the partial 2SAT logic. It is based on the standard checking tree property and the three or more long paths property.

Now we consider a special case. For convenience, we explain by the special case, but the thinking thread fits the algorithm E1 or the thinking thread can help us to understand the algorithm E1.

We have 2-unit layers and 3-unit layers. Check that the standard checking tree property holds. If the 2-unit layers contain only one useful 2SAT path, when the standard checking tree property holds, that is, when each unit in the rest three-unit layers has this useful 2SAT path, thus, there is a real long path. For convenience, suppose the 2-unit layers do not contain 1-unit layers, if they contain, because none of other units destroy any one unit in the one-unit layers, this does not affect our method and proof. Now suppose there are two part paths $s_1$ and $s_2$ over all 2-unit layers which contain utterly different units (different units, not different literals). We call these two part paths **two parallel 2SAT paths.** There may be other part paths over the 2-unit layers each of which contains some units in $s_1$ and some units in $s_2$. We call them **intersection 2SAT path** between $s_1$ and $s_2$. Now suppose there are no intersection 2SAT path between $s_1$ and $s_2$. Suppose each unit $x$ in 2-unit layers destroys (directly or indirectly destroys) at most one unit in any one 3-unit layer. If it destroys two units in a 3-unit layer, the rest one unit must not destroy any other units in 3-unit layers which has the useful unit $x$. So this does not affect our method. Then the 3-unit layers consist of three parts. Part 1: for each layer in part 1, the third unit destroys some units in $s_1$ but does not destroy units in $s_2$ and the first two units do not destroy any units in $s_1$ and $s_2$ (or, one of these two units may destroy $s_2$, but this does not affect our method). Part 2: for each layer in part 2, the third unit destroys some units in $s_2$ but does not destroy units in $s_1$ and the first two units do not destroy any units in $s_1$ and $s_2$. Part 3: for each layer in part 3, any unit does not destroy any units in $s_1$ and $s_2$. Based on the three or more long paths property, for convenience, suppose that for a part path which is not in the same one long path, there are three (old) long paths and any two units in this part path are in at least one of these three long paths. Suppose that part 1 consists of three parallel part paths $p_1$, $p_2$, $p_3$. $p_1$ contains all the first units, $p_2$ contains all the second units and $p_3$ contains all the third units (in general, $p_1$, $p_2$, $p_3$ are not three part paths, but are three unit sets). $p_3$ consists of three part paths $p_{13}$, $p_{23}$, $p_{33}$. Part 2 consists of three parallel part paths $q_1$, $q_2$, $q_3$. $q_1$ contain all the first units, $q_2$ contain all the second units and $q_3$ contain all the third units. We call this case **the special two parallel 2SAT paths case**. Continue to calculate until get a last generation. We call this structure **the two parallel 2SAT paths structure**.

Now for the special case, suppose the standard checking tree property holds. If there is no current real long path, then any one old long path must contains some third units in part 1 and some third units in part 2. If there is a current real long path which contains $s_2$, then there is one part path over part 2 which does not contain the third units in part 2 and which is in the same one long path.

Suppose that there is one long path (say *al*) which does not contain any third units in part 2 but contains $p_{13}$ and also suppose that if we delete $p_{13}$, $p_{23}$ and $p_{33}$, the standard checking tree property does not hold. Then if we only delete the first two units in some layers of $p_{13}$, the standard checking tree property must hold. Because *al* does not contain any third units in part 2, so if we delete all third units in part 2, the standard checking tree property holds. Now suppose when we only delete the first two units in some layers of $p_{33}$ (we call the unit set to contain the rest third units in these layers $r_{33}$) and also delete all the third units in part 2, the standard checking tree property holds. When it holds, there are two possible cases. In case 1, there is one (or more) real long path. In case 2, there is no real long path but based on the three or more long paths property, the standard checking tree property holds. Suppose this is the case 2, that is, $r_{33}$ is not in a current real long path. But we always have the opportunity to get a last chosen unit and its ancestor chosen units which are in $p_{13}$. We call this **the special case property**.

If there is only one two parallel 2SAT paths structure, we can easily find that the correct chosen units property holds. If there are a lot of such structures and they do not overlap, this property still holds. If some structures overlap, this does not affect that the property holds.

The correct chosen units property and the special case property are the key to prove Algorithm E1. **Understanding the special case property, the parallel part paths property, the precondition for recalculating and the correct chosen units property is the key and the hardest in the whole paper**.

Now we explain how to handle causing and caused destroying layers.

For calculating chosen units, we only consider such 3-unit layers each of which only contains one redundant unit. We call them **the 3-unit layers each with one redundant unit**. Note that **the following is the key for the 2SAT logic**: we only consider such 3-unit layers each of which only contains one redundant unit; for these 3-unit layers, the first two units (destroying parallel pair) is the first part and the third unit (redundant unit) is the second part; these two parts constitute the 2SAT. We call the 3-unit layers each of which contains two redundant units **the 3-unit layers each with two redundant units**.

If the redundant units in 3-unit layers each with one redundant unit do not contain correct units and the standard checking tree property holds, then there is at least one current real long path. This is because in 3-unit layers each with one redundant unit and in 3-unit layers without redundant units, there is a part path over these layers which contains only correct units and which does not contain those redundant units (that is, no additional contradiction pairs between these correct units), and over all other 2-unit layers, there is a part path which does not destroy this part path. So when there are no current real long paths, in the 3-unit layers each with one redundant unit, these redundant units must contain correct units. We call these **the one redundant unit property**.

For a hidden destroying parallel pair, if it does not become a visible destroying parallel pair, we do not consider it when calculating ancestor chosen units. If it becomes a visible destroying parallel pair, only note the changing father chosen unit rule. Also note **the only one time calculation of causing-caused destroying layers property** as stated above.

Note this case: a 3-unit layer contains three units *a*, *b*, and *c*, unit *x* does not destroy (exactly destroy) *y* but destroys (exactly destroys) *b* and *c*, and *a* destroys *y*, then we also say that *x* (exactly) destroys *y*. For this case, we do not need to mind other correct units affecting that *x* destroys *y*, even if *y* is a correct unit.

Now we explain the method for recalculating units.

Each time we recalculate a calculated unit for one final leaf.

A correct unit always has the opportunity to be recalculated for a correct leaf if this unit is not finished.

If just before we get the first occurred wrong 2-unit layer in algorithm E, there is only one last real long path, obviously different correct leaves' good units do not have contradictions. If such paths are more than one, the first occurred wrong 2-unit layer is only one. If we recover the third unit and delete the first two units in this layer, this third correct unit will not be changed. The next first occurred wrong 2-unit layer is also only one and is at a lower layer. So this does not affect our method.

**Note that** in this algorithm and analysis, for convenience, when we discuss how to calculate a unit *u*, and the algorithm E1 with its proof, we suppose that only the third units in 3-unit layers can be redundant units. This does not affect our method, because if two units are redundant units in a layer, the rest one must not destroy any other units in the checking tree which do not destroy *u* or which have the same one useful 2SAT path.

**For the whole paper, only the algorithm E1 and its proof are hard and in algorithm E1 the hardest is the correct chosen units property, and the special case property as well as the three or more long paths property. All others are clear and easy to understand.**

**The keys for understanding the algorithm E**: firstly understand the strong function of the standard checking tree property. This is the foundation of the algorithm. Then understand the destroying parallel pair and the additional contradiction pair caused by the destroying parallel pair. The precondition for recalculating units under a final leaf makes us not calculate 3***n* combinations. The extraordinary insight is: based on the strong function of the standard checking tree property and the destroying parallel pair, for each final leaf, we recalculate a unit at most one time and do not need to do exponential combinations. A wrong final leaf with its good units can be updated. In this way we can get the 2SAT logic. Also we always update contradiction pairs for the whole layers. All these show that this job is not based on "local consistency", but on the whole. The key is the algorithm E1.

The 3-unit layers contain two parts. The first two units are the first part and the third units (which as redundant units) are the second part. The destroying parallel pairs take up the first part. These are the base to constitute a 2SAT. As stated above, when without the separated whole redundant units case, the calculation of units has the same logic as 2SAT. When that case happens, the algorithm E1 and the $O(n)$ times ($O(n)$ times vs. one more time in 2SAT) recalculating property can handle it.

We have **key three steps for the whole algorithm** in this paper:

Key step 1: firstly get a good checking tree. In this tree, any one unit is in at least one long path. Two units which are not a contradiction pair are in at least one long path. For a contradiction pair, there is no any one long path to contain this pair. We know all contradiction pairs.

Key step 2: after the trouble units were deleted, the good checking tree is destroyed, but we repair it by the standard checking tree (see the following algorithm 1).

Key step 3: by algorithm E, we construct 2SAT based on the destroying parallel pair and the third unit which was deleted as redundant unit. Based on the standard checking tree property and the one correct last chosen unit property, for the correct final leaf, its good units are correct units and its destroying parallel pairs are wrong units and we can directly get all these, that is, they are "independent" and "absolute". For a wrong final leaf, it can be updated. These are the key for the 2SAT logic.

The key step 1 and key step 2 are easy to understand. The hardest is the key step 3, i.e., the algorithm E. In algorithm E, the hardest to understand is the algorithm E1 and the method to calculate a leaf's destroying parallel pairs and its good units. **The key for understanding them is the strong function of the standard checking tree property and the three or more long paths property, especially the partial 2SAT logic as well as the only one time calculation of causing-caused destroying layers property, the special case property and the correct chosen units property**.

In one calling of algorithm E1, there are a lot of last chosen units which have the goal generations. We call them **brother last chosen units**. If they are as final leaves, they are **brother final leaves**.

If the first (highest) final leaf's goal generation contains the first occurred wrong 2-unit layer and thus this final leaf is a correct unit, this final leaf and its correct good units will always be in 1-unit layers and will not be changed later. By the correct chosen units property, when the goal generation contains the first occurred wrong 2-unit layer, the final leaf with its good units must be correct units. If the first final leaf's goal generation does not contain the first occurred wrong 2-unit layer, at last, any new final leaf's goal generation cannot be lower than this final leaf and any one of its brother final leaves, and we cannot get the second kind goal generation under this final leaf and under its any one brother final leaf. Then we have the opportunity to get a final leaf's goal generation which is a higher generation (consider the one to one final leaf combination rule). Also for a final leaf, we calculate a unit at most one time and do not mind different final leaves above this final leaf. We call all these **the final leaf combination property**. **Note that** when the first (highest) final leaf's goal generation does not contain the first occurred wrong 2-unit layer (under the first occurred wrong 2-unit layer), no matter this leaf is correct or wrong, we may get a lot of final leaves under this final leaf in later a lot of calls of algorithm E1, but at last, we certainly can get a final leaf's goal generation higher than the first leaf's goal generation (by one to one final leaf combination rule), that is, we always can do the one to one final leaf combination, and also, for the last chosen units which are third units in hidden destroying parallel pairs' layers, we also can do the one to one final leaf combination.

A key problem is: **for different final leaves with their good units, how to get a combination of only correct final leaves whose good units are correct units in polynomial time**? By the first correct leaf property, the one to one final leaf combination rule, the correct chosen units property, and the final leaf combination property, we can get this combination in polynomial time.

### 2.3 Algorithms

Now we give a summary for the entire algorithm by the following 3 algorithms:

**Algorithm 1:** calculate a destroyed checking tree, delete the redundant units of it and get new additional contradiction pairs

Input: a destroyed checking tree which contains 2-unit layers (including 1-unit layers) $l_2$ and 3-unit layers $l_3$

Output: a standard checking tree with all layers, or the result: it cannot be a standard checking tree.

Step 1) for each unit in $l_2$, try to get a 2SAT path over $l_2$ which contains this unit. If there is no such a 2SAT path, delete this unit from $l_2$. If anyone layer in $l_2$ does not contain any units, end this algorithm and the checking tree cannot be a standard checking tree.

Step 2) for each unit (say unit $u_1$) in 3-unit layers, copy $l_2$ to $l_{21}$ and delete all units in $l_{21}$ which destroy $u_1$. Delete all units in $l_{21}$ which destroy any one unit in a 1-unit layer in $l_{21}$, until no such units. For each unit in $l_{21}$, try to get a 2SAT path over $l_{21}$ which contains this unit. If there is no such a 2SAT path, delete this unit from $l_{21}$. Now we call $l_{21}$ the unit $u_1$'s **useful 2-unit layers** and call each 2SAT path over $l_{21}$ $u_1$'s useful 2SAT path. Set $u_1$ with each of the deleted units as an additional contradiction pair (if they are not contradiction pair before). If any one layer in $l_{21}$ does not contain any units, set $u_1$'s useful units and useful unit pairs zero. Otherwise, each unit in $l_{21}$ is $u_1$'s one useful unit or one single useful unit if it is in a 1-unit layer. Any two units in a 2SAT path over $l_{21}$ is $u_1$'s one useful unit pair.

Step 3) for each unit $u$'s each useful unit or useful unit pair, if in another 3-unit layer, the units which do not destroy $u$ do not have this useful unit or useful unit pair, $u$ also has to lose this useful unit and has to lose this useful unit pair. If $u$ loses a useful unit, we delete this unit from $u$'s useful 2-unit layers and then recalculate. If $u$ loses a useful unit pair, we set this pair as a contradiction pair in $u$'s useful 2-unit layers and then recalculate. A unit's any useful unit or useful unit pair has to be in its one useful 2SAT path. So we need to recalculate the useful 2SAT paths.

Step 4) for any two units which are not contradiction pair in different 3-unit layers, if one's a single useful unit and the other one's a single useful unit are in the same layer or are a contradiction pair, set these two units as an additional contradiction pair.

Step 5) for each unit in 3-unit layers, if it destroys all units in any other one 3-unit layer, set this unit's useful units and useful unit pairs zero and set it with any other one unit as an additional contradiction pair, until no such a unit.

Step 6) if in any one 3-unit layer, all units have zero useful units, end this algorithm and the checking tree cannot be a standard checking tree.

Step 7) delete all units in 3-unit layers whose useful units are zero and move these layers to 2-unit layers $l_2$. Update the useful units and useful unit pairs of each unit in 3-unit layers on the updated 2-unit layers, that is, calculate again on the new 2-unit layers. In the mean time, if none of the three units in one layer has useful units, the checking tree cannot be a standard checking tree and end this algorithm, and if one or two units in one layer do not have useful units, move such layers to 2-unit layers and then recalculate.

Step 8) Output the standard checking tree which contains $l_2$ and $l_3$ and which contains some additional contradiction pairs. At last, each 2SAT path over $l_2$ is a useful 2SAT path. In each 3-unit layer, at least one unit has this useful 2SAT path. Each unit in $l_2$ is in at least one useful 2SAT path. Each unit in $l_3$ has useful 2SAT paths. And also the standard checking tree property holds.

Now we explain the step 3) in algorithm 1. The step 3 means that: for each unit $u$'s each useful unit, in any other one 3-unit layer, there is at least one unit which does not destroy $u$ and which also has this useful unit, and for each unit $u$'s each useful unit pair, in any other one 3-unit layer, there is at least one unit which does not destroy $u$ and which also has this useful unit pair. If not so, the unit $u$ must lose this useful unit or useful unit pair. For any two units which are a contradiction pair, we say that one destroys the other one. Though the contradiction pairs are being updated continually, we just note the current contradiction pairs we have got.

Note that all the algorithm 1 is to get such a result: to make the standard checking tree property hold. **The standard checking tree property is:** for each unit $u$'s each useful unit ($u$ is in 3-unit layers), in any other one 3-unit layer, there is at least one unit which does not destroy $u$ and which also has this useful unit, and for each unit $u$'s each useful unit pair, in any other one 3-unit layer, there is at least one unit which does not destroy $u$ and which also has this useful unit pair.

We call all deleted units in algorithm 1 the redundant units of this checking tree.

**Algorithm 2:** Call the algorithm E to calculate a standard checking tree to judge whether it contains long paths, and if it does, get one long path.

**Algorithm 3:** main algorithm

Input: a 3SAT's $m$ clauses

Output: a long path which contains $m$ literals, or an answer: no long paths

Data structures: for the main checking tree (good checking tree), remember each layer's units, remember all contradiction pairs, and also remember all utterly destroyed units which have been deleted. For the calculating main checking tree, see algorithm E.

Step 1) for layer 1, layer 2 and layer 3, that is, clause 1, clause 2 and clause 3, we calculate all the long paths. Each long path contains three units. Also calculate all indirect contradiction pairs and all utterly destroyed units. Delete utterly destroyed units and put the rest in an empty checking tree. We call it the main checking tree.

Step 2) suppose that we have got an $i$ layers' main checking tree (good checking tree) and known all indirect contradiction pairs and all utterly destroyed units for these $i$ layers. The current layer is the $i$th layer. Its three units are $a_{i1}$, $a_{i2}$, $a_{i3}$. Now we add and calculate the $i+1$th layer. Its three units are $a_{i+11}$, $a_{i+12}$, $a_{i+13}$. We add each one to the main checking tree respectively. We first add the unit $a_{i+11}$ to the main checking tree. Now the old utterly destroyed units are still utterly destroyed units and the old contradiction unit pairs are also contradiction unit pairs. We have to calculate the new utterly destroyed units and the new indirect contradiction unit pairs after the $a_{i+11}$ is added. If the checking tree does not contain such a unit which is the same variable as $a_{i+11}$ but with different sign, i.e., if $a_{i+11}$ is the variable $v$, and no other unit is the variable $-v$, then there are no new utterly destroyed units and no new contradiction unit pairs (for $a_{i+11}$). If the checking tree contains the unit $-v$, then for a new long path which contains $a_{i+11}$, any unit whose variable is $-v$ cannot be in this path.

Step 3) we want to calculate new contradiction pairs after we add a new unit to the main checking tree. Suppose the new added unit is $a_{i+11}$ and its variable is $v$. Now we copy the main checking tree to a temporary checking tree and we will delete some units in the temporary checking tree. We call this temporary checking tree the calculating main checking tree. For two units $x$ and $y$, the unit pair $x$ and $y$ is not a contradiction pair before $a_{i+11}$ is added. Now we want to know whether $x$, $y$ can be in the same long path again after we delete all the $-v$. For each unit pair which is not a contradiction pair before $a_{i+11}$ is added, we do this job one time. For the temporary checking tree, we delete the layers which contain $x$ and the layers which contain $y$. We also delete all $-x$, $-y$ and $-v$. Then we separate the layers into 3-unit layers and 2-unit layers and calculate this destroyed checking tree by the algorithm 1. If we got a standard checking tree by algorithm 1 and got a long path over the checking tree by algorithm 2, then $x$ and $y$ are not an indirect contradiction pair. Otherwise, they are. If a unit and any other one unit are a contradiction pair, it is an utterly destroyed unit. For the new added unit, each non-utterly destroyed unit (after all $-v$ are deleted) with it are a non-contradiction pair. If the new added unit's variable is $v$ and a unit whose variable is $v$ (not $-v$) is an utterly destroyed unit formerly, then we do not add this unit to the checking tree. It is also an utterly destroyed unit.

Step 4) for $a_{i+12}$ and $a_{i+13}$, we do the same job. Only when a unit pair is always an indirect contradiction pair for $a_{i+11}$, $a_{i+12}$ and $a_{i+13}$, then it is an indirect contradiction pair finally. Only when a unit is an utterly destroyed unit for $a_{i+11}$, $a_{i+12}$ and $a_{i+13}$, then it is an utterly destroyed unit finally.

Step 5) at last, after we got an $m$ layers' main checking tree and know all indirect contradiction pairs and utterly destroyed units, by the algorithm 2 we can get a long path from layer 1 to layer $m$ if exists.

Now we give a short **summary**: for $k$ layers, we know all utterly destroyed units and all indirect contradiction unit pairs. The utterly destroyed units have been deleted. At this case we can get a long path which contains $k$ units. Now we delete one or more literals (units) and each of the $k$ layers still contain at least one unit. We want to know if there are still some $k$-unit long paths. If so, we can calculate one. We first separate the $k$ layers into 2-unit layers each of which contains one or two units and 3-unit layers each of which contains three units. Then we calculate the useful units and useful unit pairs of the 3-unit layers and let the standard checking tree property hold. Then we do the algorithm E, if there are long paths, we can get one.

**Theorem 1**: the algorithm can solve the 3SAT in $O(n^6m)$.

PROOF. Suppose the number of clauses is $m$. All units are $3m$. All literals are $O(n)$. The main time is on the algorithm 1 and the algorithm E. For algorithm 1, to calculate one unit's useful units and useful unit pairs, the average time is $O(n)*O(m)$. For all units (literals), the time to calculate their useful unit pairs and useful units is $O(n)*O(m)*O(n)$ (the worst time for all literals also does not exceed this). For all the 3SAT, $O(n)$ literals (this $O(n)$ means: each time we add one layer's one unit (literal) to the checking tree) and $O(n^2)$ literal pairs. The times to do algorithm E are $O(n)*O(n^2)$. In algorithm E1, in the last generation, for each layer we calculate a last chosen unit. For each last chosen unit, for all failed final leaves and successful final leaves, for all supposed current goal generations, in each generation we have to calculate algorithm 1 one time. All last chosen units for all generations together, the times are $O(m)$. Consider that for the same literals (as chosen units) we only need to calculate one time. So in fact they together are $O(n)$ times but not $O(m)$ times. So the time complexity for the whole algorithm is: $O(n)*O(m)*O(n)*O(n)*O(n^2)*O(n)= O(n^6m)$. This is the whole time complexity. Other jobs are not bigger than the time.

## 3 HOW TO UNDERSTAND THIS PAPER?

In a whole, understanding the **key three steps for the whole algorithm** is the key. The hardest is the key step 3 and the algorithm E1. The key for the polynomial is: the partial 2SAT logic, the final leaf combination property and the precondition for recalculating units under a final leaf make us not need to do 3**$n$ combinations.

In detail, understand the following:

A 3SAT contains $n$ variables and $m$ clauses. Each clause contains three literals. We call the literals $x$ and $-x$ a direct contradiction pair and say that $x$ destroys $-x$ or $-x$ destroys $x$. We call a solution of the 3SAT a long path. A long path contains $m$ literals and each one in each clause. Any two literals in a long path cannot be a direct contradiction pair. For two different literals, if there is no any one long path to contain them, we call them an indirect contradiction pair (if they are not a direct contradiction pair) and we say that one destroys the other one indirectly. For a literal, if there is no any one long path to contain it, we call it an utterly destroyed literal or unit. Suppose that we have known all the indirect contradiction pairs and all the utterly destroyed literals. So we call such a problem a solved 3SAT problem, or a good checking tree. Now, we delete one or more literals from this 3SAT (we call the deleted literals **trouble units**). We call this problem a destroyed solved 3SAT problem, or a destroyed checking tree. It still keeps $m$ layers. So if a long path for it exists, this long path still contains $m$ literals (may repeat). We want to recalculate all the indirect contradiction pairs and utterly destroyed literals after these literals are deleted. Please note: the old utterly destroyed literals are still utterly destroyed literals and the old indirect contradiction pairs are still indirect contradiction pairs. Solving this problem can be as the basis to solve 3SAT. In this paper, the key is that we develop a polynomial time algorithm for solving a destroyed solved 3SAT. Then based on this, our algorithm solves 3SAT.

How to solve a destroyed solved 3SAT? This is the key for the entire algorithm. Suppose for $k$ layers, we have known all the indirect contradiction pairs and utterly destroyed units (or literals). Now we add a new layer $a_1$, $a_2$, $a_3$. At first we only add $a_1$. Suppose that $x_1$, $x_2$ are not a contradiction pair for the $k$ layers. For each such a pair, we have to calculate. Now we want to know whether it is a contradiction pair after adding $a_1$. We still only consider the $k$ layers. We delete the **trouble units** $-a_1$, $-x_1$, $-x_2$ in the $k$ layers. Now we separate the $k$ layers into 2-unit layers (a layer may contain only two or one unit) and 3-unit layers. If we can get a long path (contains $k$ units) in them, then $x_1$, $x_2$ is still not a contradiction pair, otherwise, it is a contradiction pair. This job is polynomial.

Understand the above 3 algorithms.

We take an example to explain. We have nine layers: -1, 2, 3 (that is, $-x_1$, $x_2$, $x_3$, they are three literals); 1, -2, 3; 1, 2, -3; -1, -2, -3; -4, 5, 6; 4, -5, 6; 4, 5, -6; -4, -5, -6; -3, 6, 7. We first consider the first three layers. If only considering the three layers, 2, 3, 1 is a long path and there are other long paths. A long path contains three units coming from three different layers. So 1, 2 is not an indirect contradiction pair. For a unit pair, if only there is no long path to contain this pair, it is an indirect contradiction pair (if not direct). Also for one unit, if only there is no long path to contain it, it is an utterly destroyed unit. So for these three layers, there are no utterly destroyed units and no indirect contradiction pairs. Now we add the layer -1, -2, -3. We first add -1. For this -1, each long path contains four units and it must contain -1 in the fourth layer. We still consider three layers but delete 1 (temporarily delete) in the three layers. So 1, 2; 1, 3; 1, -3; 1, -2 are indirect contradiction pairs. We add -2, -3 in this way. At last, if a pair is an indirect contradiction pair in all the three cases, it is an indirect contradiction pair for the four layers. So for the four layers, there are still no utterly destroyed units and no indirect contradiction pairs. Suppose for the 9 layers, we know all indirect contradiction pairs and utterly destroyed units. Now we add a new layer 3, -6, 7 (i.e., $x_3$, $-x_6$, $x_7$). Each time add one unit. We first add 3. Now an old contradiction pair is still a contradiction pair. But an old non-contradiction pair may become an indirect contradiction pair. For each non-contradiction pair, we have to calculate. 5, 6 is not an old contradiction pair. Now we delete -3, -5 and -6 in the nine layers. We get the 2-unit layers: 1, 2; -1, -2; 4, 6; 4, 5; -4 (one unit in this layer); 6, 7; six layers. We also get 3-unit layers: -1, 2, 3; 1, -2, 3; -4, 5, 6; three layers. Now for each unit in the 3-unit layers, we calculate its useful 2SAT path when it is together with the six 2-unit layers. For example, we calculate -1's useful 2SAT path. We temporarily delete 1 (also delete the units which indirectly destroy -1. If then some 2-unit layers do not contain units, -1 does not have useful units) in the six 2-unit layers, then calculate 2SAT paths in the 6 layers. Each unit in such a 2SATpath are -1's one useful unit. We do not remember all 2SAT paths, but remember -1's all useful units and useful unit pairs. In this way, we can calculate the useful units and useful unit pairs of each unit in the 3-unit layers. For each unit *u*'s each useful unit or useful unit pair, if in another 3-unit layer, the units which do not destroy *u* do not have this useful unit, *u* also has to lose this useful unit or useful unit pair. All useful units or useful unit pairs of the three units in a layer are this layer's useful units or useful unit pairs. For any two 3-unit layers, their useful units or and useful unit pairs have to be the same. If we can get such a long path, then 5, 6 is still a non contradiction pair. Only when a pair is a contradiction pair for the unit 3, and also for -6, and for 7, then it is a new contradiction pair.

Please note the concept "utterly destroyed unit" in the paper about page 2. We only delete utterly destroyed units permanently. We keep all other units for the whole process. When we want to add a unit *x*, if *x* as a literal has been in former layers and is an utterly destroyed unit (the same literal) formerly, we do not have to add *x*. Usually the indirect contradiction pairs become more and more. If all the unit pairs are indirect or direct contradiction pairs, this mean: no sat.